\documentclass[12pt]{article}
\usepackage[top=2.5cm, bottom=2.5cm, left=2.5cm, right=2.5cm]{geometry} 

\usepackage{amsmath}
\usepackage{graphicx}		
\usepackage{amsfonts}		
\usepackage{amssymb}		
\usepackage{amscd}			
\usepackage{slashed}
\usepackage{hyperref}		
\usepackage{mathtools}

\usepackage[toc,page]{appendix}	
\numberwithin{equation}{section}
\hypersetup{
    pdftitle={5d Higgs Branch Localization, Seiberg-Witten Equations and Contact Geometry},
    pdfauthor={Yiwen Pan},    		
    pdfnewwindow=true,      		
    colorlinks=true,       			
    linkcolor=blue,          		
    citecolor= blue,        		
    filecolor= black,      			
    urlcolor= black           		
}

\begin{document} 

\title{5d Higgs Branch Localization, Seiberg-Witten Equations and Contact Geometry}
\date{}
\author{Yiwen Pan\footnote{Email address:  \href{mailto:yiwen.pan@stonybrook.edu}{yiwen.pan@stonybrook.edu}}
\\
\textit{C. N. Yang Institute for Theoretical Physics}\\
\textit{Stony Brook University, Stony Brook, NY 11794}}

\maketitle
\begin{abstract}
\noindent In this paper we apply the idea of Higgs branch localization to 5d supersymmetric theories of vector multiplet and hypermultiplets, obtained as the rigid limit of $\mathcal{N} = 1$ supergravity with all auxiliary fields. On supersymmetric K-contact/Sasakian background, the Higgs branch BPS equations can be interpreted as 5d generalizations of the Seiberg-Witten equations. We discuss the properties and local behavior of the solutions near closed Reeb orbits. For $U(1)$ gauge theories, which can be straight-forwardly generalized to theories whose gauge group can be completely broken, we show the suppression of the deformed Coulomb branch, and the partition function is dominated by 5d Seiberg-Witten solutions. For squashed $S^5$ and $Y^{pq}$ manifolds, we show the matching between poles in the perturbative Coulomb branch matrix model, and the bound on local winding numbers of the BPS solutions.  \end{abstract}

\thispagestyle{empty}	

\newpage
\tableofcontents

\section{Introduction}

	Starting from \cite{Pestun:2007rz}, there had been numerous development in exactly computing quantities in supersymmetric field theories on curved manifolds, using the localization method. Intuitively, these developments can be loosely classified into two approaches, which should be the two sides of a whole but not fully understood story. 

	One side of the story mostly concerns the exactly computable quantities of theories defined on selected interesting geometries. These developments allow us to study the fine structures of these quantities.

	In 3-dimension, progresses have been made to understand the structures of the supersymmetric partition functions on selected geometries. In particular, the supersymmetric partition function on squashed $S^3$ (smooth homological 3-spheres) is shown to be expressed in terms of double-sine functions \cite{Hama:2011ea, Imamura:2011wg, Alday:2013lba, Nian:2013qwa}. Multiple-sine functions are a family of interesting functions that enjoy factorization properties. Indeed, these properties are studied in \cite{Pasquetti:2011fj}; later the $S^3$ partition function, originally written as a matrix model, is unpacked into a product of vortex-anti-vortex partition functions \cite{Shadchin:aa}. This result later inspired the Higgs branch localization in 3-dimension \cite{Fujitsuka:2013aa}\cite{Benini:2013yva}. See also the Higgs branch localization on $S^3 \times S^1$ \cite{Peelaers:2014ima}.

	In 5-dimension, there are also similar results. Starting from the round spheres $S^5$ \cite{Kallen:2012fk, Kim:2012aa}, then on the squashed $S^5$ \cite{Imamura:2012aa, Kim:2012uq, Lockhart:2012aa}, and later on the Sasaki-Einstein manifolds \cite{Qiu:2013ab, Qiu:2014aa,Schmude:2014lfa}, the perturbative partition functions were computed, and the full non-perturbative partition functions were conjectured. Supersymmetric theories and their partition functions on other type of manifolds are also studied in detail \cite{Kim:2012ab, Kim:2013,Kim:2014kta}. Similar to 3-dimensional theories, the 5d perturbative results are expressed in terms of a matrix model with triple-sine functions (or their certain products) as integrand. As a member of the multiple-sine functions family, triple-sine function also has a similar factorization property: it factorizes into three pieces (two pieces when $d = 3$) corresponding to three closed polar Reeb orbits (two polar orbits when $d = 3$), which leads to the conjecture of the full non-perturbative partition function in Coulomb branch on Sasaki-Einstein manifolds \cite{Qiu:2013ab, Qiu:2014aa}.

	It is therefore natural to believe that the Higgs branch localization in 3-dimension can be generalized to 5-dimension, as a manifestation of the factorization property. Namely, factorizing the matrix model and performing the contour integral should pick up residues from the poles, and these residues are the local contributions from some new BPS solutions localized to certain loci on the manifold.

	There is another side of the story which concerns more about general geometric structures that support supersymmetries. Following the idea of obtaining supersymmetry on a curved manifold by taking rigid limit of suitable supergravity \cite{Festuccia:2011aa}, various developments took place to understand the relation between supersymmetry and the underlying geometries \cite{Alday:2013lba}\cite{Dumitrescu:2012ha, Dumitrescu:2012ly, Klare:2012gn, Cassani:2012ri, Closset:2012aa,Cassani:2012ri,Klare:2013dka,Closset:2014pda}. In particular, it is shown in \cite{Closset:2013vra} that $\mathcal{N} = 2$ and $\mathcal{N} = 1$ partition functions in 3d and 4d are holomorphic function of transversally holomorphic foliation moduli and complex structure moduli.

	The above two approaches should somehow be consistent. For instance, it would be interesting to ask the questions like ``can we start from a general supersymmetric theory on a 3-manifold as in \cite{Closset:2012aa} and carry out the Higgs branch localization'', or ``what geometric structures do the ingredients of the matrix model, or the vortex/anti-vortex partition functions actually correspond to, if the whole partition function is an invariant or holomorphic function of certain geometric structures''. At the moment, to the best of the author's knowledge, these kinds of questions are not fully understood.

	Therefore, in this note, we try to start from general backgrounds that support 5d supersymmetry and apply the idea of Higgs branch localization in this general setting. We find that the Higgs branch localization locus can be interpreted as a 5d generalization of perturbed Seiberg-Witten equations on symplectic or K\"{a}hler 4-manifolds. If we recall that solutions to (a sequence of) perturbed Seiberg-Witten equations are related to pseudo-holomorphic curves in symplectic 4-manifolds\cite{Taubes1996}, the 5d partition functions can be viewed in some sense as invariants that captures ``pseudo-holomorphic'' objects in contact manifolds.

	The content of this note will be organized as follows:

	1. In section \ref{section-1}, we start from 5d $\mathcal{N} = 1$ supergravity and review some geometric implications of the existence of supersymmetry. In particular, we study the generalized Killing spinor equation and show how it is related to K-contact geometry.

	2. In section \ref{section-2}, we write down the general supersymmetric theories of $\mathcal{N} = 1$ vector multiplet and hypermultiplets, which can be obtained by taking rigid limit. Then we redefine the field variables, and obtain corresponding cohomological complexes. By adding $Q$-exact terms we obtain the new BPS equations. On a K-contact background, the Higgs branch BPS equations can be interpreted as generalization of Seiberg-Witten equations, by introducing the generalized Tanaka-Webster connection:
	\begin{equation}
		\left\{ \begin{gathered}
  		F_a^{d\kappa } = \frac{1}{2}\left( {\zeta  - {{\left| \alpha  \right|}^2} + {{\left| \beta  \right|}^2}} \right)d\kappa  \hfill \\
  		F_a^{02} = 2i\bar \alpha \beta  \hfill \\ 
		\end{gathered}  \right.\;,\;\;\;\;\left\{ \begin{gathered}
  		{{\bar \partial }_a}\alpha  + \bar \partial _a^*\beta  = 0 \hfill \\
  		\mathcal{L}_R^a\alpha  = \mathcal{L}_R^a\beta  = 0 \hfill \\ 
		\end{gathered}  \right.
	\end{equation}

	We will show that Sasakian structures are concrete examples where solutions to the above equations have simple behavior. We also extend the discussion to more general K-contact backgrounds, and study the local behavior of solutions around closed Reeb orbits.

	3. In section \ref{section-3}, We show that as the Higgs branch parameter $\zeta \to + \infty$, one can suppress the deformed Coulomb branch if the matter content and the Chern-Simons level satisfy a certain inequality. We also show that on squashed $S^5$ and $Y^{pq}$ manifolds, the bound on local winding numbers of Higgs branch BPS solutions corresponds to poles in the Coulomb branch matrix model integrand. To do so, we interpret the shift of the form ${\Sigma _i}{\omega _i}/2$ in the 1-loop determinant as the the $R$-component of the ``Chern-connection'' on the anti-canonical line bundle of the K-contact structure.

	4. In the appendix, we summarize useful aspects of contact geometry and a review of $\operatorname{Spin}^\mathbb{C}$-structures on any contact metric manifolds. The generalized Tanaka-Webster connection and its Dirac operator are also reviewed, which are closely related to the BPS equations.

\section{From Supergravity to K-contact Geometry \label{section-1}}

	\subsection{Symplectic-Majorana Spinors, Self-duality and Chirality\label{section-1.1}}

	In this subsection we will discuss properties of symplectic-Majorana spinors and their bilinears on 5-dimensional manifolds.

	\vspace{20pt}

	\underline{\emph{Symplectic-Majorana spinors}}

	A symplectic-Majorana spinor $\xi_I$ with $I = 1,2$ satisfies
	\begin{equation}
		\overline {\xi _I^\alpha }  = {\epsilon ^{IJ}}{C_{\alpha \beta }}\xi _J^\beta ,
	\end{equation}
where $C$ is the charge conjugation matrix $C = C_+$, $C_{\alpha\beta} = - C_{\beta\alpha}$. We define two products $(,)$ and $\left\langle , \right\rangle $ for any two symplectic-Majorana spinors $\xi_I$ and $\chi_J$ as
	\begin{equation}
		\left( {{\xi _I}{\chi _J}} \right) \equiv \sum\limits_{\alpha ,\beta  = 1,2} {\xi _I^\alpha {C_{\alpha \beta }}\chi _J^\beta } ,\;\;\;\;\;\;\;\;\left\langle {\xi ,\chi } \right\rangle  \equiv {\epsilon ^{IJ}}\left( {{\xi _I}{\chi _J}} \right).
	\end{equation}
Note that the $(,)$ is anti-symmetric, while $\left\langle , \right\rangle $ is symmetric and positive semi-definite. We also denote the action of any differential $k$-form $\omega$ on any spinor $\psi$ by
	\begin{equation}
		\omega  \cdot \psi  \equiv \frac{1}{{k!}}{\omega _{{m_1}...{m_k}}}{\Gamma ^{{m_1}...{m_k}}}\psi.
	\end{equation}

	\vspace{10pt}
	\underline{\emph {Bilinears of a symplectic-Majorana Spinor}}

	Given any spinor $\xi$, one can define several bi-linears using the products defined above.
	\begin{itemize}
		\item Real scalar $s \equiv \left\langle {\xi ,\xi } \right\rangle > 0$. This is the norm-squared of the spinor $\xi$.
		\item Real vector $R^m \equiv - \left\langle {\xi ,{\Gamma ^m}\xi } \right\rangle $\footnote{The minus sign is conventional; changing the sign will swap ``self-duality'' and ``chirality'' discussed later.}. The norm-squared of $R$ is ${R^m}{R_m} = {s^2}$, or equivalently $\iota_R \kappa = s^2$, where we define the metric-dual 1-form $\kappa_m = g_{mn} R^n$.
		\item Several 2-forms ${\left( {{\Theta _{IJ}}} \right)_{mn}} \equiv \left( {{\xi _I}{\Gamma _{mn}}{\xi _J}} \right)$.
	\end{itemize}
These bilinears satisfy various algebraic identities following from the Fierz identities, which are summarized in the Appendix [\ref{appendix-a}].

	\vspace{20pt}
	\underline{\emph{5-dimensional Self-duality}}

	Given any nowhere-vanishing spinor $\xi$, we construct the associated set of quantities $(s, R, \kappa, \Theta_{IJ})$. By rescaling we set $s = 1$. We then use them to decompose any $p$-forms
	\begin{equation}
		\omega  = \kappa  \wedge {\iota _R}\omega  + {\iota _R}\left( {\kappa  \wedge \omega } \right) \equiv {\omega _V} + {\omega _H},
	\end{equation}
and we call $\omega_H$ ($\omega_V$ respectively) is called the horizontal\footnote{Note that $\iota_R \omega_H = 0$ is the characteristic feature of a horizontal form $\omega_H$.} (vertical) part of $\omega$.  We then decompose the space of $p$-forms ${\Omega ^p}\left( M \right) = \Omega _V^p\left( M \right) \oplus \Omega _H^p\left( M \right)$, and define the projection operators ${\pi _H} \equiv {\iota _R}\circ \kappa  \wedge ,\;\;{\pi _V} \equiv \kappa  \wedge {\iota _R}$.

	Similarly we decompose $TM = T{M_V} \oplus T{M_H}$ such that $\kappa \left( {\forall X \in T{M_H}} \right) = 0$.

	Let $*$ be the Hodge star operator of metric $g$. Then we have operator ${\iota _R}*:{\Omega ^p}\left( M \right) \to {\Omega_H ^{4 - p}}\left( M \right)$, such that
	\begin{equation}
		{\left( {{\iota _R}*} \right)^2} = \pi_H.
	\end{equation}
In view of this, we can restrict ${{\iota _R}*}$ onto $\Omega _H^2\left( M \right)$ and decompose horizontal 2-forms into self-dual $(+)$ and anti-self-dual 2-forms $(-)$, according to the eigenvalues of $\iota_R *$:
	\begin{equation}
		{\iota _R}*{\omega ^ \pm } =  \pm {\omega ^ \pm },\;\;\;\;\forall \omega^\pm  \in {\Omega ^ \pm }\left( M \right) \subset {\Omega _H}\left( M \right).
	\end{equation}
So the final result is one can decompose any 2-form $\omega$ into
	\begin{equation}
		\omega  = {\omega _V} + {\omega ^ + } + {\omega ^ - }, \;\;\forall \omega  \in {\Omega ^2}\left( M \right) = \Omega _V^2\left( M \right) \oplus {\Omega ^ + }\left( M \right) \oplus {\Omega ^ - }\left( M \right).
	\end{equation}

	Before moving to next subsection, we remark that following from Fierz-identities, the 2-forms $\Theta_{IJ}$ are always horizontal self-dual:
	\begin{equation}
		{\iota _R}*{\Theta _{IJ}} = {\Theta _{IJ}}.
	\end{equation}
	Also one can straight-forwardly extend the self-duality to the general case where $s \ne 1$.
	
	Another remark is that any anti-self-dual 2-form $\omega^-$ annihilates $\xi_I$ (the very $\xi_I$ used to define $R^m$):
	\begin{equation}
		\omega _{mn}^ - {\Gamma ^{mn}}{\xi _I} = 0, \;\;\;\;\forall \omega^- \in \Omega^-(M).
	\end{equation}

	\vspace{10pt}
	\underline{\emph{Chirality}}

	As reviewed in appendix [\ref{appendix-a}], we define the chiral operator $\Gamma_C \equiv - R^m \Gamma_m$, which satisfies chirality (following from Fierz-identities (\ref{Fierz}) and the assumption $s = 1$)
	\begin{equation}
		{\Gamma _C}{\xi _I} = {\xi _I}
	\end{equation}
Naturally, $\Gamma_C$ induces a decomposition of spinor bundle $S = S_+ \oplus S_-$, and we denote the projection operators
	\begin{equation}
		{P_ \pm } \equiv \frac{1}{2}\left( {1 \pm {\Gamma _C}} \right):S \to {S_ \pm }.
	\end{equation}

	\subsection{5-dimensional $\mathcal{N} = 1$ Minimal Off-shell Supergravity}

	In this subsection we briefly review 5-dimensional minimal off-shell supergravity discussed in \cite{Zucker:2000uq}\cite{Zucker:1999fn}\cite{Kugo:fk} (see also literatures on superspace formalism \cite{Kuzenko:2014eqa}\cite{Kuzenko:2013rna}), and then extract the generalized Killing spinor equation by taking the rigid limit, following the idea of \cite{Festuccia:2011ve}.

	The Weyl multiplet contains the following bosonic field content (note that there is a curly $\mathcal{V}$ and straight $V$)
	\begin{equation}
		\mathcal{G}_\text{Boson} = \left\{ {e_m^A,\;\;\;\;{\mathcal{A}_m},\;\;\;\;{\mathcal{V}_{mn}},\;\;\;\;{t_{IJ}},\;\;\;\;C,\;\;\;\;{{\left( {{V_m}} \right)}_{IJ}}} \right\}.
	\end{equation}
Here $I, J = 1,2$ are indices of $SU(2)_\mathcal{R}$ symmetry, $\mathcal{A}_m$ is the abelian gauge field corresponding to central charge with field strength $\mathcal{F} = d \mathcal{A}$, $\mathcal{V}$ is a 2-form, $C$ is a scalar. Field $t_{IJ}$ and $V_{IJ}$ are both $SU(2)_\mathcal{R}$ triplet, meaning that
	\begin{equation}
		\overline {{t_{IJ}}}  = {\epsilon ^{IK}}{\epsilon ^{JL}}{t_{KL}}.
	\end{equation}
and similarly for $V_{IJ}$. The fermionic field content contains
	\begin{equation}
		{\mathcal{G}_{{\text{Fermion}}}} = \left\{ {\psi_I ,\;\;\;\;\eta_I } \right\},
	\end{equation}
where $\psi$ is the gravitino, $\eta$ is the dilatino. Finally, the supergravity transformation $\delta_\text{Sugra}$ has symplectic-Majorana parameter $\xi_I$.
	
	To obtain a supersymmetric theory of some matter multiplet on some manifold $M$, one can first couple it to the above Weyl multiplet $\mathcal{G}$, and then set all fields in $\mathcal{G}$ to some background values that is invariant under the supergravity transformation $\delta_\text{Sugra}$. In particular, we set the fermions $(\psi, \eta)$ to zero background, and requires two spinorial differential equations (with coefficients comprised with fields $\{V, \mathcal{V}, \mathcal{F}, t_{IJ}, C\}$)
	\begin{equation}
		{\delta _{{\text{Sugra}}}}\psi  = 0,\;\;\;\;\;\;\;\;{\delta _{{\text{Sugra}}}}\eta  = 0,
	\end{equation}
with transformation parameter $\xi_I$, and look for background values of $\{V, \mathcal{V}, \mathcal{F}, t_{IJ}, C\}$ that admit a solution $\xi_I$. The result of such procedure is \cite{Festuccia:2011aa,Dumitrescu:2012ly,Dumitrescu:2012ha,Klare:2012gn}:
	\begin{itemize}
		\item Supersymmetry transformation $Q$ obtained from $\delta_\text{Sugra}$ by substituting in background values of $\{V, \mathcal{V}, \mathcal{F}, t_{IJ}, C\}$.
		\item A $Q$-invariant Lagrangian from the coupled supergravity Lagrangian, where all remaining bosonic fields from $\mathcal{G}$ are auxiliary background fields.
		\item Some geometric data, including metric $g$, $p$-forms and so forth, determined by combinations of $\{V, \mathcal{V}, \mathcal{F}, t_{IJ}, C\}$.
	\end{itemize}
 
	First of all, we focus on the equation $\delta_\text{Sugra} \psi = 0$, which we refer to as the \emph{generalized Killing spinor equation} in the following discussion. The generalized Killing spinor equation reads
	\begin{equation}
		{\nabla _m}{\xi _I} = {t_I}^J{\Gamma _m}{\xi _J} + {\mathcal{F}_{mn}}{\Gamma ^n}{\xi _I} + \frac{1}{2}{\mathcal{V}^{pq}}{\Gamma _{mpq}}{\xi _I}
		\label{Killing-eq-0},
	\end{equation}
where $\nabla$ contains the usual Levi-Civita spin connection as well as $SU(2)_\mathcal{R}$ gauge field $V_m$ when acting on objects with $I,J$ indices. Strictly speaking, $\xi_I$ is a section of the bundle $S \otimes V$ where $V$ is a $SU(2)_\mathcal{R}$-vector bundle on which ${(V_M)_I}^J$ is defined, therefore we should require $M$ to be a spin manifold.

	Equation (\ref{Killing-eq-0}) is studied in \cite{Pan:2013uoa}, where geometric restrictions imposed by different numbers of solutions is discussed. Subsequently, in \cite{Imamura:2014aa} both differential equations $\delta \psi = \delta \eta = 0$ are solved in a coordinates patch. It is shown that, locally,  deformations of auxiliary fields that preserves (\ref{Killing-eq-0}) and (\ref{dilatino-eq}) can be realized as $Q$-exact deformation or gauge transformations. This suggests that path integrals of appropriate observables may be topological or geometrical invariants. For us, it is important to note that $\delta_\text{Sugra}\eta = 0$ implies (which we may call the \emph{dilatino equation})
	\begin{equation}
		 4\left( {{\nabla _m}{t_I}^J} \right){\Gamma ^m}{\xi _J} + 4{\nabla _m}{\mathcal{V}^{mn}}{\Gamma _n}{\xi _I} + 4{t_I}^J\left( {{\mathcal{F}_{mn}} + 2{\mathcal{V}_{mn}}} \right){\Gamma ^{mn}}{\xi _J} + {\mathcal{F}_{mn}}{\mathcal{F}_{kl}}{\Gamma ^{mnkl}}{\xi _I} =  - C{\xi _I}
		 \label{dilatino-eq}
	\end{equation}
This will be used to ensure the closure of the rigid $\mathcal{N} = 1$ supersymmetry. Note that the field $C$ can be solved using this equation in terms of $\{V, \mathcal{F}, \mathcal{V}, {t_I}^J\}$, by contracting both sides with $\xi^I$:
	\begin{equation}
		4{R_n}{\nabla _m}{\mathcal{V}^{mn}} - 4{\left( {\mathcal{F} + 2\mathcal{V}} \right)_{mn}}{\left( {{t^{IJ}}{\Theta _{IJ}}} \right)^{mn}} + 2{\left( {{\iota _R}*\mathcal{F}} \right)^{mn}}{\mathcal{F}_{mn}} = sC
		\label{C}
	\end{equation}
where $R$, $\Theta$ and $s$ are defined using $\xi_I$ as explained earlier.

So to summarize, for the rigid limit to give rise to a rigid supersymmetry, we are required to study the Killing spinor equations and the dilatino equation
	\begin{equation*}
		\left\{ \begin{gathered}
  		{\nabla _m}{\xi _I} = {t_I}^J{\Gamma _m}{\xi _J} + {\mathcal{F}_{mn}}{\Gamma ^n}{\xi _I} + \frac{1}{2}{\mathcal{V}^{pq}}{\Gamma _{mpq}}{\xi _I} \hfill \\[0.5em]
  		4\left( {{\nabla _m}{t_I}^J} \right){\Gamma ^m}{\xi _J} + 4{\nabla _m}{\mathcal{V}^{mn}}{\Gamma _n}{\xi _I} + 4{t_I}^J\left( {{\mathcal{F}_{mn}} + 2{\mathcal{V}_{mn}}} \right){\Gamma ^{mn}}{\xi _J} + {\mathcal{F}_{mn}}{\mathcal{F}_{kl}}{\Gamma ^{mnkl}}{\xi _I} =  - C{\xi _I} \hfill \\ 
		\end{gathered}  \right.
	\end{equation*}
where one can immediately solve $C$ in terms of other auxiliary fields using (\ref{C}).

	\subsection{Generalized Killing Spinor Equation}

	In this subsection we will review some basic properties the Killing spinor equations that are relevant to later discussions. Some terminology in K-contact geometry will be reviewed in the following subsection.

	As introduced in the previous subsection, the Killing spinor equation for symplectic-Majorana spinor $\xi_I$ is
	\begin{equation}
	\boxed{
		{\nabla _m}{\xi _I} = {t_I}^J{\Gamma _m}{\xi _J} + {\mathcal{F}_{mn}}{\Gamma ^n}{\xi _I} + \frac{1}{2}{\mathcal{V}^{pq}}{\Gamma _{mpq}}{\xi _I}
		\label{Killing-eq}
	}\;.
	\end{equation}
Recall that we have several background fields coming from the Weyl multiplet: $\mathcal{F}$ is a closed 2-form, and $\mathcal{V}$ is a usual 2-form as the field strength of $\mathcal{A}$, ${t_I}^J$ is a triplet of scalars. The connection $\nabla$ contains the Levi-Civita spin connection and possibly a non-zero $SU(2)_\mathcal{R}$ background gauge field $V_m$ acting on the $I$-indices. All these fields are from the Weyl multiplet $\mathcal{G}$ and we call them \emph{auxiliary fields} below.

	Equation (\ref{Killing-eq}) can also be written in a more convenient form
	\begin{equation}
	\boxed{
		{\nabla _m}{\xi _I} = {\Gamma _m}{\tilde \xi _I} + \frac{1}{2}{\mathcal{P}^{pq}}{\Gamma _{mpq}}{\xi _I}, \;\;\;\;{{\tilde \xi }_I} \equiv {t_I}^J{\xi _J} + \frac{1}{2}{\mathcal{F}_{mn}}{\Gamma ^{mn}}{\xi _I},\;\;\;\;\mathcal{P} \equiv \mathcal{V} - \mathcal{F}
	}\;.
	\label{Killing-eq-con}
	\end{equation}

	\vspace{10pt}
	\underline{\emph{1. Symmetries}}

	The Killing spinor equation enjoys several symmetries that will help simplify later discussions.
	\begin{itemize}
		\item Background $SU(2)_\mathcal{R}$ symmetry, which acts on the $I$-index.
		\item Shifting symmetry: one can shift the auxiliary fields $\mathcal{F}$ and $\mathcal{V}$ by any anti-self-dual\footnote{Defined using $R^m \equiv - (\xi_I \Gamma^m \xi^I)$, and in the sense of general $s$ as we remarked earlier.} 2-form $\Omega^-$
			\begin{equation}
				\mathcal{F} \to \mathcal{F} + {\Omega ^ - },\;\;\mathcal{V} \to \mathcal{V} + {\Omega ^ - }.
			\end{equation}
		and the equation is invariant.
		\item Other symmetries related to the many degrees of freedoms discussed in \cite{Imamura:2014aa}. We will come back to this shortly.
	\end{itemize}

	\vspace{10pt}
	\underline{\emph{2. Solving the Killing spinor equation}}

	Let $\xi_I$ be a solution to the Killing spinor equation (\ref{Killing-eq}). Then one can construct bi-linears $s$, $R^m$, $\kappa_m$ and $\Theta_{IJ}$ using $\xi_I$. By directly applying equation (\ref{Killing-eq}), one obtains several differential properties of these bi-linears:
	\begin{itemize}
		\item ${\nabla _m}s = 2{R^n}{\mathcal{F}_{nm}} \Leftrightarrow ds = 2{\iota _R}\mathcal{F}$ and therefore ${\mathcal{L}_R}s = 0,\;\;{\mathcal{L}_R}\mathcal{F} = 0$, where we have used the Bianchi identity $d\mathcal{F} = 0$.
		\item ${\nabla _m}{R_n} =  2{t^{IJ}}{\left( {{\Theta _{IJ}}} \right)_{mn}} - 2s{\mathcal{F}_{mn}} - 2{\left( {{\iota _R}*\mathcal{V}} \right)_{mn}}$, or equivalently,
		\begin{equation}
			d\kappa  = 4\left( {{t^{IJ}}{\Theta _{IJ}}} \right) - 4s\mathcal{F} - 4{\iota _R}*\mathcal{V}, \;\;\;\;\;\;	\mathcal{L}_Rg = 0.
		\end{equation}
	\end{itemize}

	Using the above basic properties, one can partially solve
	\begin{equation}
		\mathcal{F} =  - \frac{{d\kappa }}{{4s}} - \frac{{{\Omega ^ + } + {\Omega ^ - }}}{s},\;\;\;\;{\mathcal{V}_H} = {s^{ - 1}}\left( {{t^{IJ}}{\Theta _{IJ}} + {\Omega ^ + } - {\Omega ^ - }} \right).
		\label{aux-fields-sol}
	\end{equation}
Recall that the Killing spinor equation enjoys a shifting symmetry, and therefore one can always set $\Omega^- = 0$ in the above solutions; so let us do this. Then we have
	\begin{equation}
		s\left( {{\mathcal{F}_H} + {\mathcal{V}_H}} \right) =  - \frac{{d\kappa_H }}{4} + {t^{IJ}}{\Theta _{IJ}}
		\label{FV}
	\end{equation}

	To further simplify later discussion, let us apply the results in \cite{Imamura:2014aa}. The Killing spinor equation and the dilatino equation are solved locally, and it is shown that the auxiliary fields are highly unconstrained by the existence of solutions. 

	The freedom can be understood by looking at the Fierz identities. In some sense, solving the equations is just to properly match the ``$\Gamma$-matrices structure'' in (\ref{Killing-eq}) and (\ref{dilatino-eq}). Note that one can use the Fierz-identities
	\begin{equation}
		- \frac{1}{{4s}}{\lambda ^{KL}}{\left( {{\Theta _{KL}}} \right)_{mn}}{\Gamma ^{mn}}{\xi _I} = {\lambda _I}^J{\xi _J},\;\;\;\;{\lambda ^{KL}}{\left( {{\Theta _{KL}}} \right)_{mn}}{\Gamma ^n}{\xi _I} =  - {\lambda _I}^J\left( {{R_m} + s{\Gamma _m}} \right){\xi _J}
	\end{equation}
to alter the $\Gamma$-structures. Hence one can adjust the $SU(2)_\mathcal{R}$-gauge field $(V_m)_{IJ}$ to cancel terms with $\Gamma$-matrices in (\ref{Killing-eq}), and consequently other auxiliary fields are left unconstrained.
	
	We can use the local freedom in $s$ and $t_{IJ}$ to smoothly adjust them such that $s = 1$ and $\operatorname{tr}(t^2) \equiv {t_I}^J {t_J}^I = -1/2$ in a patch. Note that given a global Killing spinor solution, $s$ and $\operatorname{tr}(t^2)$ should be patch-independent functions, and therefore, the adjustment can be made global. Therefore, let us deform the solution and auxiliary fields such that \emph{globally} $s \equiv 1 \Rightarrow \iota_R \mathcal{F} = 0$ and $\operatorname{tr}(t^2) \equiv - 1/2$. Furthermore, it is shown in \cite{Imamura:2014aa} that resulting deformations in the actions are $Q$-exact, and therefore the above adjustment does not change the expectation values of BPS observables.

	\vspace{10pt}
	\underline{\emph{3. A special class of solutions}}

	Equation (\ref{FV}) implies that it is interesting to look at a special class of solutions where the auxiliary fields $\mathcal{F}$ and $\mathcal{V}$ are such that
	\begin{equation}
		\left( {\mathcal{F} + {\mathcal{V}_H}} \right) = \Lambda d\kappa  \Rightarrow d\kappa  = \frac{4}{{\Lambda  + 1}}{t^{IJ}}{\Theta _{IJ}},\;\;\;\;{\iota _R}\mathcal{F} = 0.
		\label{special-class}
	\end{equation}
for some constant $\Lambda \in \mathbb{R}$. This implies $\kappa$ is a contact 1-form, namely it satisfies (assuming $t_{IJ} \ne 0$)
	\begin{equation}
		\kappa  \wedge d\kappa  \wedge d\kappa \;\; \propto \;\; \kappa  \wedge \left( {{t^{IJ}}{\Theta _{IJ}}} \right) \wedge \left( {{t^{IJ}}{\Theta _{IJ}}} \right) \ne 0.
	\end{equation}

	\vspace{10pt}
	\underline{\emph{4. Towards a K-contact structure}}

	Now the bi-linears from the special class of solutions satisfy various conditions:
	\begin{equation}
		\left\{ \begin{gathered}
  		\kappa  \wedge d\kappa  \wedge d\kappa  \ne 0,\;\;\;\;\kappa \left( R \right) = 1,\;\;{\iota _R}d\kappa  = 0 \hfill \\[0.5em]
  		{\left( {d\kappa } \right)_{mn}} = \frac{4}{1 + \Lambda}{\left( {t\Theta } \right)_{mn}},\;\;\;\;{\mathcal{L}_R}g = 0,\;\;\;\;{\kappa _m} = {g_{mn}}{R^n} \hfill \\ 
		\end{gathered}  \right..
	\end{equation}
The first row tells us that $(\kappa, R)$ defines a contact structure, while the second row implies the contact structure closely resembles a K-contact structure. The only \emph{violation} appears in
	\begin{equation}
		d\kappa  = \frac{4}{{1 + \Lambda }}{\left( {t\Theta } \right)_{mn}} = \left[ {\frac{1}{{1 + \Lambda }}} \right]\left( {2{g_{mk}}{\Phi ^k}_n} \right),\;\;\;\;{\Phi ^m}_k{\Phi ^k}_n =  - \delta _n^m + {R^m}{\kappa _n}.
	\end{equation}
where we defined $\Phi  = 2\left( {{t^{IJ}}{\Theta _{IJ}}} \right)$, instead of the standard form
	\begin{equation}
		{\left( {d\kappa } \right)_{mn}} = 2{g_{mk}}{\Phi ^k}_n,\;\;\;\;{\Phi ^m}_k{\Phi ^k}_n =  - \delta _n^m + {R^m}{\kappa _n}
	\end{equation}
	
	It is easy to bring the system to a standard \emph{K-contact structure}. Let us use an adapted veilbein $\{e^A\}$ such that
	\begin{equation}
		g = \sum\limits_a {{e^a}{e^a}}  + \kappa  \otimes \kappa, \;\;\;\;{e^5} = \kappa ,\;\;\;\;{\iota _R}{e^{a = 1,2,3,4}} = 0,\;\;\;\;\Phi(e^1) = e^2, \Phi(e^3) = e^4.
	\end{equation}
Define a function $\lambda$ by ${\lambda ^2} \equiv {\left( {1 + \Lambda } \right)^{ - 1}}$, and we rescale the horizontal piece of $g$ by $g \to  g' = \sum\limits_a {{{e'}^a}{{e'}^a}}  + \kappa  \otimes \kappa $ with ${{e'}^a} = \lambda {e^a}$.

	With the new metric, the quantities $(\kappa, R, g', \Phi)$ defines a standard K-contact structure on $M$:
	\begin{equation}
		\left\{ \begin{gathered}
  		\kappa  \wedge d\kappa  \wedge d\kappa  \ne 0,\;\;\;\;\kappa \left( R \right) = 1,\;\;{\iota _R}d\kappa  = 0 \hfill \\[0.5em]
  		{\left( {d\kappa } \right)_{mn}} = 2g'_{mk} {\Phi^k}_n,\;\;\;\;{\mathcal{L}_R}g' = 0,\;\;\;\;{\kappa _m} = {g_{mn}}{R^n} \hfill \\ 
		\end{gathered}  \right.
	\end{equation}

	Along with the change in metric, one needs to properly deform the auxiliary fields to preserve the equation (\ref{Killing-eq}). By explicitly working out the change in spin connection ${\omega^A}_B$, one can identify the required deformations in $\mathcal{F}$ and $\mathcal{V}$ (both are deformed by multiples of $d\kappa$), which indeed also preserve the condition (\ref{special-class}), and therefore no inconsistency arises. Finally, since the deformed auxiliary fields are independent and unconstrained as shown in \cite{Imamura:2014aa}, the resulting deformations preserves the two equations (\ref{Killing-eq}) and (\ref{dilatino-eq}) (and field $C$ can be solved using (\ref{C})),  and the actions  are deformed by $Q$-exact, hence do not change the expectation values of BPS observables.

	To summarize, any solution to (\ref{Killing-eq}) of the special class can be transformed into a standard one, such that the resulting set of geometric quantities $(\kappa, R, g, \Phi)$ form a K-contact structure. Later we will discuss BPS equations on K-contact and Sasakian backgrounds, where the equations are better behaved than on completely general supersymmetric backgrounds.

	\subsection{K-contact Geometry}

	In this subsection, we summarize most important aspects and formula of contact geometry that we will frequently use in later discussions. For more detail introduction, readers may refer to appendix [\ref{appendix-c}].

	\vspace{10pt}
	\underline{1.\emph{ Contact structure}}

	A contact structure is most conveniently described in terms of a contact 1-form. A contact 1-form on a $2n + 1$-manifold is a 1-form $\kappa$ such that
	\begin{equation}
		\kappa  \wedge {\left( {d\kappa } \right)^n} \ne 0.
	\end{equation}
This is analogous to the definition of a symplectic form on an even dimensional manifold.

	We can associate quantities ($R, g, \Phi$) to $\kappa$ called a \emph{contact metric structure}, such that
	\begin{equation}
		{\kappa _m}{R^m} = 1,\;\;\;\;{R^m}d{\kappa _{mn}} = 0,\;\;\;\;{\Phi ^m}_k{\Phi ^k}_n =  - \delta _n^m + {R^m}{\kappa _n},\;\;\;\;{\left( {d\kappa } \right)_{mn}} = 2{g_{mk}}{\Phi ^k}_n
	\end{equation}
The vector field $R$ is called the \emph{Reeb} vector field, and $\Phi$ is like an almost complex structure in directions orthogonal to $R$.

	On a contact metric 5-manifold, we will frequently use an \emph{adapted vielbein} $\{e^A\}$, $\{e_A\}$, such that ${e_5} = R,\;\;\Phi \left( {{e_1}} \right) = {e_2},\;\;\Phi \left( {{e_3}} \right) = {e_4}$, and
	\begin{equation}
		d\kappa  = 2\left( {{e^1} \wedge {e^2} + {e^3} \wedge {e^4}} \right),\;\;\;\;g = \sum\limits_{a = 1,2,3,4} {{e^a} \otimes {e^a}}  + \kappa  \otimes \kappa ,
	\end{equation}
Note that the first equation implies $d\kappa$ is self-dual, namely ${\iota _R}*d\kappa  = d\kappa$. We will also use the complexification of $\{e^A\}$:
	\begin{equation}
		\left\{ \begin{gathered}
  		{e^{{z_i}}} \equiv {e^{2i - 1}} + i{e^{2i}},\;\;\;\;{e^{{{\bar z}_i}}} \equiv {e^{2i - 1}} - i{e^{2i}},\;\;\;\;{e^5} = \kappa  \hfill \\
  		{e_{{z_i}}} \equiv \frac{1}{2}\left( {{e_{2i - 1}} - i{e_{2i}}} \right),\;\;{e_{{{\bar z}_i}}} \equiv \frac{1}{2}\left( {{e_{2i - 1}} + i{e_{2i}}} \right),\;\;\;\;{e_5} = R \hfill \\ 
		\end{gathered}  \right.
	\end{equation}
so that $\displaystyle \left\{ {1,\frac{1}{{\sqrt 2 }}{e^{{{\bar z}_1}}},\frac{1}{{\sqrt 2 }}{e^{{{\bar z}_2}}},\frac{1}{2}{e^{{{\bar z}_1}}} \wedge {e^{{{\bar z}_2}}}} \right\}$ are orthonormal.

	\vspace{20pt}
	\underline{2. \emph{K-contact and Sasakian structure}}

	A \emph{K-contact structure} is a contact structure $\kappa$ and the associated $(R, g, \Phi)$, such that
	\begin{equation}
		\mathcal{L}_R g= 0 \;\;\;\; \Leftrightarrow \;\;\;\; \nabla_m R_n + \nabla_n R_m = 0
	\end{equation}
Note that one immediately has $\mathcal{L}_R \Phi = 0$.

	For a general contact structure, the integral curves of $R$, or equivalently, the 1-parameter diffeomorphisms $\varphi_R(t)$ (the \emph{Reeb flow}) generated by $R$, can have three types of behavior. The \emph{regular} or \emph{quasi-regular} types are such that the flow are free or semi-free $U(1)$ action, respectively. The \emph{irregular} type is such that the flow is not $U(1)$, and therefore the integral curves of $R$ generally are \emph{not} closed orbits. 

	Generic irregular Reeb flows are difficult to study, however, situation can be improved when the contact structure is K-contact. In this case, the \emph{closure} of the Reeb flow (it preserves $g$ by definition), viewed as a subgroup of the $\operatorname{Isom}(M,g)$, is a torus $T^k \subset \operatorname{Isom}(M,g)$; $k$ is called the \emph{rank} of the K-contact structure. On a K-contact 5-manifold, $1 \le k \le 3$.

	Finally, a \emph{Sasakian structure} is a K-contact structure with additional property
	\begin{equation}
		{\nabla _m}{\Phi ^k}_n = {g_{mn}}{R^k} - {\kappa _n}\delta _m^k
	\end{equation}
Sasakian structures are the K\"{a}hler structures in the odd-dimensional world. They satisfies certain integrability condition, and all quantities discussed above, as well as some metric connections associated with $g$, live in great harmony. We will later see that on Sasakian structures, the Higgs branch BPS equations have very simple behavior, very much like Seiberg-Witten equations on K\"{a}hler manifolds.

	To end this section, we tabulate the correspondence between the structures (including some we haven't mentioned) in even and odd dimensional worlds.
	\begin{center}
	\begin{tabular}{ c | c }
  		Even & Odd \\
  		\hline
  		Symplectic  & Contact \\
  		Almost Hermitian & K-contact \\
  		Complex & Cauchy-Riemann \\
  		K\"{a}hler & Sasakian\\
  		K\"{a}hler-Einstein & Sasaki-Einstein \\
  		HyperK\"{a}hler & 3-Sasakian
	\end{tabular}
	\end{center}

\section{Higgs Branch Localization and 5d Seiberg-Witten Equation \label{section-2}}

	In this section, we begin by reviewing the 5-dimensional $\mathcal{N} = 1$ vector multiplet and hypermultiplet. Then we consider deforming the theory with $Q$-exact terms to localize the path-integral. We discuss the deformed Coulomb branch solutions and the Higgs branch. We rewrite the Higgs branch equations and interpret them as 5-dimensional generalizations of Seiberg-Witten equations on symplectic 4-manifolds. We also discuss basic properties of solutions to the 5d Seiberg-Witten equations, including their local behavior near closed Reeb orbits.

	\subsection{Vector-multiplet and Hyper-multiplet}

	~~~~\underline{1. \emph{Vector-multiplet}}
	
	The Grassman odd transformation $Q$ of vector multiplet $(A_m, \sigma, \lambda_I, D_{IJ})$ can be obtained directly from $\mathcal{N} = 1$ supersymmetry transformation, which can be obtained by taking the rigid limit of coupled supergravity in \cite{Kugo:2000}. Using a symplectic-Majorana spinor $\xi_I$ satisfying Killing spinor equation (\ref{Killing-eq-con}), the transformation can be written as
	\begin{equation}
		\left\{ \begin{array}{l}
		{Q	 }{A_m} = i{\epsilon ^{IJ}}\left( {{\xi _I}{\Gamma _m}{\lambda _J}} \right)\\[0.5em]
		{Q }\sigma  = i{\epsilon ^{IJ}}\left( {{\xi _I}{\lambda _J}} \right)\\[0.5em]
		\displaystyle{Q }{\lambda _I} =  - \frac{1}{2}{F_{mn}}{\Gamma ^{mn}}{\xi _I} + \left( {{D_m}\sigma } \right){\Gamma ^m}{\xi _I} + {D_I}^J{\xi _J} + 2\sigma {\tilde \xi _I} \\[0.5em]
		\displaystyle {Q }{D_{IJ}} =  - i\left( {{\xi _I}{\Gamma ^m}{D_m}{\lambda _J}} \right) + \left[ {\sigma ,\left( {{\xi _I}{\lambda _J}} \right)} \right] + i({\tilde \xi _I}{\lambda _J}) - \frac{i}{2} \mathcal{P}_{mn} (\xi_I \Gamma^{mn} \lambda_J) + \left( {I \leftrightarrow J} \right)
		\end{array} \right.,
		\label{vector brst}
	\end{equation}
where ${D_m}\left(  \cdot  \right) = {\nabla _m} - i\left[ {{A_m}, \cdot } \right]$, and $\tilde \xi_I$ is defined in (\ref{Killing-eq-con}). Here the spinor $\xi_I$ is Grassman even. The transformation squares to
	\begin{equation}
		{Q ^2} =  - i{\mathcal{L}^A_R} + {\mathcal{G}_{s\sigma }} + {\mathcal{R}_{{R_I}^J}} + {L_\Lambda }，
	\end{equation}
where $\mathcal{G}$ is gauge transformation, $\mathcal{R}$ is $SU(2)_\mathcal{R}$ rotation acting on a generic field $X_I$ as ${\mathcal{R}_{{R_I}^J}}{X _I} = {R_I}^J{X _J}$, and $L$ is Lorentz rotation acting on spinors. The parameters are
	\begin{equation}
		\left\{ \begin{gathered}
  		{R^m} =  - ({\xi _I}{\Gamma ^m}{\xi ^I}) \hfill \\
  		s = ({\xi _I}{\xi ^I}) \hfill \\ 
		\end{gathered}  \right.,\;\;\;\;\left\{ \begin{gathered}
  		{\Lambda _{mn}} = \left( { - 2i} \right)\left( {({\xi _J}{\Gamma _{mn}}{{\tilde \xi }^J}) - s\left( {\mathcal{P}_{mn}^ +  - \mathcal{P}_{mn}^ - } \right)} \right) \hfill \\
  		{R_I}^J = 2i\left[ {3({\xi _I}{{\tilde \xi }^J}) + {\mathcal{P}^{mn}}{{\left( {{\Theta _I}^J} \right)}_{mn}}} \right] \hfill \\ 
		\end{gathered}  \right.
	\end{equation}
and we used the vector field $R^m$ to define self-duality $\Omega_H^\pm (M)$, see section \ref{section-1.1}.

Note that, similar to \cite{Hosomichi:2012fk}, there is a term in $\delta^2 D_{IJ}$ that breaks the closure of the supersymmetry algebra, of the form
	\begin{equation}
		{\delta ^2}{D_{IJ}} = ... + \sigma \left[ {({\xi _I}{\Gamma ^m}{\nabla _m}{{\tilde \xi }_J}) + \frac{1}{2}{\mathcal{P}^{mn}}({\xi _I}{\Gamma _{mn}}{{\tilde \xi }_J}) + \left( {I \leftrightarrow J} \right)} \right].
		\label{violation}
	\end{equation}
Such a term vanishes if there exists a function $u$ and a vector field $v_m$ such that
	\begin{equation}
		\slashed{\nabla}{\tilde \xi _I} + \frac{1}{2}{\mathcal{P}_{mn}}{\Gamma ^{mn}}{\tilde \xi _I} =  u {\xi _I} + {v_m}{\Gamma ^m}{\xi _I}
	\end{equation}
In the case of $\mathcal{P}_{mn} = 0$, one can show that $v = 0$ and the function $u$ always exists and is proportional to the scalar curvature of the metric $(g, \nabla^\text{LC})$. In the presence of $\mathcal{P}_{mn}$, by explicitly expanding every term, one can show that
	\begin{equation}
		\begin{gathered}
  		\;\;\;\;\slashed{\nabla}{{\tilde \xi }_I} + \frac{1}{2}{\mathcal{P}^{mn}}{\Gamma _{mn}}{{\tilde \xi }_I} \hfill \\
   		= \left( {{\nabla _m}{t_I}^J} \right){\Gamma ^m}{\xi _J} + {\nabla _m}{\mathcal{V}^{mn}}{\Gamma _n}{\xi _I} + t\left( {{\mathcal{F}_{mn}} + 2{\mathcal{V}_{mn}}} \right){\Gamma ^{mn}}{\xi _I} + \frac{1}{4}{\mathcal{F}_{kl}}{\mathcal{F}_{mn}}{\Gamma ^{mnkl}}\xi  \hfill \\
  		\;\;\;\; + \frac{3}{2}{\mathcal{F}_{mn}}{\mathcal{F}^{mn}}\xi  - 2{\mathcal{F}_{mn}}{\mathcal{V}^{mn}}\xi  + 5( {{t_I}{{^K}_K}^J} ){\xi _J} - {\nabla _m}\left( {{\mathcal{V}^{mn}} - {\mathcal{F}^{mn}}} \right){\Gamma _n}\xi.  \hfill \\ 
		\end{gathered} 
	\end{equation}
We observe that the first row is just the left hand side of (\ref{dilatino-eq}), and therefore, recalling $( {{t_I}{{^K}_K}^J} ){\xi _J} = 1/2\left( {{t_L}{^K}{t_K}^L} \right){\xi _I}$,
	\begin{equation}
		\begin{gathered}
  		\;\;\;\;{\slashed{\nabla}}{{\tilde \xi }_I} + \frac{1}{2}{\mathcal{P}^{mn}}{\Gamma _{mn}}{{\tilde \xi }_I} \hfill \\[0.5em]
   		= \left[ {\frac{5}{2}\left( {{t_L}^K{t_K}^L} \right) - \frac{1}{4}C + \frac{3}{2}{\mathcal{F}_{mn}}{\mathcal{F}^{mn}} - 2{\mathcal{F}_{mn}}{\mathcal{V}^{mn}}} \right]{\xi _I} - {\nabla _m}\left( {{\mathcal{V}^{mn}} - {\mathcal{F}^{mn}}} \right){\Gamma _n}\xi . \hfill \\ 
		\end{gathered} 
	\end{equation}
Namely, we found the required function and the vector field to be
	\begin{equation}
		\left\{ \begin{gathered}
  		u = \frac{5}{2}\left( {{t_L}^K{t_K}^L} \right) - \frac{1}{4}C + \frac{3}{2}{\mathcal{F}_{mn}}{\mathcal{F}^{mn}} - 2{\mathcal{F}_{mn}}{\mathcal{V}^{mn}} \hfill \\
  		{v^n} = {\nabla _m}\left( {{\mathcal{F}^{mn}} - {\mathcal{V}^{mn}}} \right), \hfill \\ 
		\end{gathered}  \right.
	\end{equation}
We therefore confirmed that the term (\ref{violation}) vanishes happily, thanks to (\ref{dilatino-eq}). Finally, we point out that function $u$ will appear in the supersymmetric Yang-Mills Lagrangian for the vector multiplet (which is denoted as $P$ in \cite{Imamura:2014aa}), in the form of
	\begin{equation}
		{\mathcal{L}_{{\text{YM}}}} = ... - 4u{\sigma ^2} + 4i\sigma {F_{mn}}{\mathcal{P}^{mn}} - {\mathcal{P}_{mn}}\left( {{\lambda _I}{\Gamma ^{mn}}{\lambda ^I}} \right).
	\end{equation}

	\vspace{10pt}
	\underline{2. \emph{Hypermultiplet}}

	A hypermultiplet in 5-dimension consists of a set of scalars $\phi _I^A$, two spinors $\psi^A$ and a set of auxiliary scalars $\Xi_{I'}^A$. Here $I, I' = 1,2$ are two \emph{different} copies of $SU(2)$ indices (in particular, $I$ corresponds to the $SU(2)_\mathcal{R}$-symmetry), while $A = 1,2$ is a separate $Sp(1)$ index. They satisfy reality conditions
	\begin{equation}
		\overline {\phi _I^A}  = {\epsilon ^{IJ}}{\Omega _{AB}}\phi _J^B, \;\;\;\;\overline {{\psi ^{A\alpha }}}  = {\Omega _{AB}}{C_{\alpha \beta }}{\psi ^{B\beta }},\;\;\;\;\overline {\Xi_{I'}^A}  = {\Omega _{AB}}{\epsilon ^{I'J'}}\Xi_{J'}^B.
	\end{equation}
In the above, $\Omega_{AB}$ is the invariant $Sp(1)$ tensor $\Omega_{12} = - \Omega_{21} = 1$.
		
	The reality conditions reduces the independent components. The field $\phi_I^A$ can be represented by two complex scalar $\phi^{1,2}$
	\begin{equation}
		\phi _{I = 1}^A = \frac{1}{{\sqrt 2 }}\left( {\begin{array}{*{20}{c}}
  		{{\phi ^1}} \\ 
  		{{\phi ^2}} 
		\end{array}} \right),\;\;\;\;\phi _{I = 2}^A = \frac{1}{{\sqrt 2 }}\left( {\begin{array}{*{20}{c}}
  		{ - \overline {{\phi ^2}} } \\ 
  		{\overline {{\phi ^1}} } 
		\end{array}} \right)
	\end{equation}
and similarly for the field $\Xi_{I'}^A$. The field $\psi^A$ can be represented in terms of one spinor $\psi$
	\begin{equation}
		{\psi ^A} = \left( {\begin{array}{*{20}{c}}
  		\psi  \\ 
  		{ - C \bar \psi } 
		\end{array}} \right)
	\end{equation}

	In the following, we couple the hypermultiplet to a $U(N_c)$ vector multiplet by setting the independent fields to be in appropriate representation of $U(N)$, for instance,
	\begin{equation}
		{\phi ^ 1 }:{\rm N},\;\;\;\;\;\;\;\;{\phi ^ 2 }:\bar {\rm N},\;\;\;\;\;\;\;\;\psi : {\rm N}, \;\;\;\;\;\;\;\;\bar \psi :\bar N
	\end{equation}
We define $D_m$ on any field $\Phi$ in hypermultiplet as ${D_m}\Phi  = {\nabla _m}\Phi  - i{A_m}\left( \Phi  \right)$, where $\nabla_m$ may contain spin connection and $SU(2)_\mathcal{R}$-the background gauge field ${(V_m)}_{IJ}$.

	It is well-known that one cannot write down an off-shell supersymmetry transformation for a hypermultiplet with finitely many auxiliary fields. But it is possible to write down a Grassmann odd transformation $Q$ which squares to bosonic symmetries. As transformation parameters, we use a symplectic-Majorana spinor $\xi_I$ satisfying Killing spinor equation (\ref{Killing-eq-con}), and an additional $SU(2)'$-symplectic-Majorana spinor $\hat \xi_{I'}$, satisfying
	\begin{equation}
		({\hat \xi _I}{\hat \xi ^I}) = \left( {{\xi _I}{\xi ^I}} \right) = s,\;\;\left( {{\xi _I}{\Gamma ^m}{\xi ^I}} \right) = - {R^m} =  - ({\hat \xi _I}{\Gamma ^m}{\hat \xi ^I}),\;\;({\hat \xi _{I'}}{\xi _J}) = 0.
	\end{equation}
One can view $\hat \xi_{I'}$ as a orthogonal complement of $\xi_I$ in the spinor space, and therefore corresponds to anti-chiral spinors, in the sense that ${\Gamma _C}{\xi _I} = s {\xi _I}$, ${\Gamma _C}{{\hat \xi }_{I'}} =  - s {{\hat \xi }_{I'}}$ where $\Gamma_C \equiv - R^m \Gamma_m$. Using the Fierz identities, one can show completeness relations for an arbitrary spinor $\varsigma$ (see section \ref{section-1.1} and appendix [\ref{appendix-a}]):
	\begin{equation}
		{\xi _I}\left( {{\xi ^I}\varsigma } \right) =  - \frac{1}{4}\left( {s + {\Gamma _C}} \right)\varsigma \xrightarrow{{s = 1}} - \frac{1}{2}{P_ + }\varsigma ,\;\;\;\;{{\hat \xi }_{I'}}({{\hat \xi }^{I'}}\varsigma ) =  - \frac{1}{4}\left( {s - {\Gamma _C}} \right)\varsigma \xrightarrow{{s = 1}} - \frac{1}{2}{P_ - }\varsigma .
		\label{completeness}
	\end{equation}

	The Grassman odd transformation $Q$ is as follows:
	\begin{equation}
		\left\{ \begin{gathered}
  		Q \phi _I^A =  - 2i\left( {{\xi _I}{\psi ^A}} \right) \hfill \\[0.5em]
  		Q {\psi ^A} = {\epsilon ^{IJ}}{\Gamma
  		 ^m}{\xi _I}{D _m}\phi _J^A + i \epsilon^{IJ}\xi_I\sigma \phi_J^A - 3{{\tilde \xi }^I}\phi _I^A + {\mathcal{P}_{pq}}{\epsilon ^{IJ}}{\Gamma ^{pq}}{\xi _I}\phi _J^A + {\epsilon ^{I'J'}}{{\hat \xi }_{I'}}{\Xi_{J'}} \hfill \\[0.5em]
  		Q {\Xi_{J'}}^A = 2{{\hat \xi }_{J'}}\left( {i{\Gamma ^m}{D _m}{\psi ^A} + \sigma \psi^A + \epsilon^{KL} \lambda_K \phi_L^A - \frac{i}{2}{\mathcal{P}_{pq}}{\Gamma ^{pq}}{\psi ^A}} \right) \hfill \\ 
		\end{gathered}  \right..
		\label{hypersusy}
	\end{equation}

	The transformation squares to the bosonic symmetries
	\begin{equation}
		{Q ^2} =  - i{\mathcal{L}^A_R} + {\mathcal{G}_{s\sigma }} + {\mathcal{R}_{{R_I}^J}} + {\mathcal{R}_{{{\hat R}_{I'}}^{\;\;J'}}} + {L_\Lambda }.
	\end{equation}
where $\mathcal{G}$ is the gauge transformation, $\mathcal{R}$ is $SU(2)$ rotations on $I, J$ and $I', J'$ indices, $L$ is Lorentz rotation; the parameters are
	\begin{equation}
		\left\{ \begin{gathered}
  		{\Lambda _{mn}} = \left( { - 2i} \right)\left( {( {{\xi _J}{\Gamma _{mn}}{{\tilde \xi }^J}} ) - s{{\left( {\mathcal{P}_{mn}^ +  - \mathcal{P}_{mn}^ - } \right)}}} \right) \hfill \\[0.5em]
  		{R_I}^J = 2i\left[ {3({\xi _I}{{\tilde \xi }^J}) + {\mathcal{P}^{mn}}{{\left( {{\Theta _I}^J} \right)}_{mn}}} \right] \hfill \\[0.5em]
  		{{\hat R}_{I'}}^{\;\;J'} = \left( { - 2i} \right)\left[ {\left( {{{\hat \xi }_{I'}}{\Gamma ^m}{\nabla _m}{{\hat \xi }^{J'}}} \right) - \frac{1}{2}{\mathcal{P}_{mn}}\left( {{{\hat \xi }_{I'}}{\Gamma ^{mn}}{{\hat \xi }^{J'}}} \right)} \right] \hfill \\ 
		\end{gathered}  \right..
	\end{equation}
As in previous sections we define the function $s \equiv (\xi_I \xi^I)$, and $\Omega_H^\pm(M)$ is defined with respect to the vector field $R^m \equiv - (\xi_I \Gamma \xi^I)$.

	\subsection{Twisting, $Q$-exact Deformations and Localization Locus}

	In this subsection, we first review a redefinition (the twisting) of field variables in vector multiplet and hypermultiplet. Then using the redefined variables, we introduce the $Q$-exact deformation terms and derive the localization locus. Here we explicitly used gauge group $U(N_c)$, but in general one can choose gauge groups with $U(1)$-components.

	\vspace{20pt}
	\underline{\emph{The twisting}}

	First introduced in \cite{Kallen:2012fk}\cite{Kallen:2012zr} in the context of Sasaki-Einstein backgrounds, all field variables with $I$ or $I'$ indices can be ``twisted'' (invertible using Fierz-identities (\ref{Fierz})) using $\xi_I$ and $\hat \xi_{I'}$. In our situation, assuming $s = 1$ and recalling (\ref{aux-fields-sol}), we define:
	\begin{equation}
		\left\{ \begin{gathered}
  		{\Psi _m} \equiv \left( {{\xi _I}{\Gamma _m}{\lambda ^I}} \right),\;\;\;\;{\chi _{mn}} \equiv \left( {{\xi _I}{\Gamma _{mn}}{\lambda ^I}} \right) + \left( {{\kappa _m}{\Psi _n} - {\kappa _n}{\Psi _m}} \right) \hfill \\[0.5em]
  		H = 2{F_A^ + } + {D^{IJ}}{\Theta _{IJ}} + \sigma \left( {2{t^{IJ}}{\Theta _{IJ}} + d{\kappa ^ + } + 4{\Omega ^ + }} \right) \hfill \\ 
		\end{gathered}  \right.,\;\;
		\left\{ \begin{gathered}
  		\phi _ + ^A \equiv {\epsilon ^{IJ}}{\xi _I}\phi _J^A \hfill \\
  		\Xi_ - ^A \equiv {\epsilon ^{I'J'}}{{\hat \xi }_{I'}}\Xi_{J'}^A \hfill \\ 
		\end{gathered}  \right.
	\end{equation}
After such redefinitions, $\chi$ and $H$ are both horizontal self-dual two forms with respect to vector field $R^m$, $\phi^A_+$ are chiral spinors\footnote{More explicitly, with the gauge index in place,
	\begin{equation}
		{\left( {\phi _ + ^{A = 1}} \right)^a} = {\xi ^1}{\left( {{\phi ^{A = 1}}} \right)^a} + {\xi ^2}{( - \overline {{\phi ^{A = 2}}} )^a}
	\end{equation}.} while $\Xi^{A =1,2}_-$ are anti-chiral.

	In terms of these twisted field variables, the originally complicated BRST transformations can be rewritten into very simple forms:
	\begin{equation}
		\left\{ \begin{gathered}
  		QA = i\Psi  \hfill \\
  		Q\sigma  =  - i{\iota _R}\Psi  \hfill \\
  		Q\Psi  =  - {\iota _R}F_A + {d_A}\sigma  \hfill \\
  		Q\chi  = H  \hfill \\
  		QH =  - i{\mathcal{L}^A_R}\chi  - \left[ {\sigma ,\chi } \right]  \hfill \\ 
		\end{gathered}  \right., \;\left\{ \begin{gathered}
  		Q \phi _ + ^A = i{P_ + }{\psi ^A} \hfill \\
  		Q {\psi ^A} = \slashed{D}\phi _ + ^A + i\sigma \phi _ + ^A + \frac{1}{8}{(d\kappa) _{mn}}{\Gamma ^{mn}}\phi _ + ^A + \Xi_ - ^A \hfill \\ 
  		Q {\Xi_-^A} =  - i{P_ - }\slashed{D}{\psi ^A} - \sigma {P_ - }{\psi ^A} - {\Psi ^m}\left( {{\Gamma _m} + {R_m}} \right){\phi_+ ^A}\\
		\end{gathered}  \right..
	\end{equation}
In order to derive $Q\psi^A$ and $Q\Xi_-^A$, one needs to use the symmetry $( {{\xi _I}{{\tilde \xi }_J}} ) = ( {{\xi _J}{{\tilde \xi }_I}} )$ and completeness relations (\ref{completeness}). Also we will use $d\kappa \cdot \phi_+ \equiv 1/2 (d\kappa)_{mn}\Gamma^{mn} \phi_+$ to simplify the notations in the following discussions.

	For later convenience, we separate $Q \psi$ into chiral and anti-chiral part:
	\begin{equation}
		Q \psi _ + ^A = {P_ + }\slashed{D}\phi _ + ^A + i\sigma \phi _ + ^A + \frac{1}{4}d\kappa  \cdot \phi _ + ^A,\;\;\;\;\;\;\;\; 	Q \psi _ - ^A = {P_ - }\slashed{D}\phi _ + ^A + \Xi_ - ^A,
	\end{equation}
which implies that
	\begin{equation}
		{Q^2} =  - i\left( {{R^m D_m} + \frac{1}{4}d\kappa  \cdot } \right) - \sigma 
	\end{equation}

Note that $d\kappa$ is horizontal, and therefore its Clifford multiplication does not change chirality, similar to that in 4-dimension. Also, the new spinorial variables have reality condition, for instance, where $C$ is the charge conjugation matrix,
	\begin{equation}
		\overline {\phi _ + ^A}  = {\Omega _{AB}}C\phi _ + ^B
	\end{equation}

	\vspace{10pt}
	\underline{\emph{$Q$-exact terms}}

	We are now ready to introduce the $Q$-exact terms. There are three of them\footnote{In the second line, expanding the terms and using the reality, one obtains, for instance the kinetic term ${D_m}{\phi ^{A = 1,a}}{D^m}\overline {{\phi ^{A = 1}_a}}  + {D_m}{\phi ^{A = 2}_a}{D_m}\overline {{\phi ^{A = 2,a}}} $, where $a$ is the gauge index that were suppressed.}
	\begin{equation}
		\left\{ \begin{gathered}
  		Q{V_{{\text{Vect}}}} = Q\int {\operatorname{Tr} \left( {\chi  \wedge *\left( {2F_A^ +  - H} \right) + \frac{1}{2}\Psi  \wedge *Q\bar \Psi } \right)}  \hfill \\
  		Q{V_{{\text{Hyper}}}} = Q\int_M {{\Omega _{AB}}} \overline {Q{\psi ^A}} {\psi ^B} \hfill \\
  		Q{V_{{\text{Mixed}}}} = Q\int_M {{\text{Tr}}\left[ {2\chi  \wedge *h\left( {{\phi _ + }} \right)} \right]}   \hfill \\ 
		\end{gathered}  \right.\;\;,
	\end{equation}
where $h$ maps the ``spinor'' $\phi_+^A$ in the hypermultiplet to a \emph{adjoint-valued} self-dual 2-form $h(\phi_+)$. Its explicit form will be given in
	\begin{equation}
		h\left( \phi  \right) = \alpha \left( \phi  \right) - \frac{\zeta}{2} d{\kappa ^ + } - F_{A_0/2}^+,
	\end{equation}
where $\zeta \sim \zeta {1_{{N_c} \times {N_c}}}$ is a ``fake'' FI-parameter taking value in the $\mathfrak{u}(1)$-component of the Lie-algebra $\mathfrak{u}(N_c)$, $A_0$ is a non-dynamical gauge field which we put in \emph{by hand} for later computations, taking value in the $\mathfrak{u}(1)$ in $\mathfrak{u}(N_c)$ with the property $\iota_R F_{A_0/2} = 0$ ($F_{A_0/2} = 1/2 dA_0 $)\footnote{It is straight-forward to generalize to other gauge groups with $U(1)$ components generated by $h_a$. There one picks $\zeta  = {\zeta ^a}{h_a}$, and $A_0$ takes value in the diagonal $h_1$ proportional to identity. For gauge groups without any $U(1)$-components, one cannot perform the Higgs branch localization described in this article.}. $\alpha$ is an \emph{adjoint-valued} bilinear map from chiral spinors to self-dual 2-forms, whose explicit form will be given in a spinor basis later on, schematically of the form
	\begin{equation}
		{\alpha _{mn}}{\left( \phi  \right)^a}_b = (\phi _ + ^{A = 1,a}{\Gamma _{mn}}\overline {\phi _{ + ,b}^{A = 1}} ),
	\end{equation}
Up to this point, other than $s = 1$, we make no assumption on the background geometry. Hence $d\kappa$ does not have to be self-dual; $d\kappa^+$ means we extract the self-dual part from $d \kappa$. To ensure positivity, we need to analytically continue $\sigma  \to - i\sigma $, $\Xi_-^A \to i\Xi_-^A$.

	Now one can expand all terms, and integrate out auxiliary field $H$, or equivalently, impose the field equation of $H$:
	\begin{equation}
		H = {F_A^ + } + h\left( \phi  \right).
	\end{equation}
Then the bosonic $Q$-exact terms reads
	\begin{equation}
		{\left( {F_A^ +  + h({\phi _ + })} \right)^2} + \frac{1}{2}{\left( {{\iota _R}{F_A}} \right)^2} + {\left( {{d_A}\sigma } \right)^2} + {\left| {{D_A}{\phi _ + } + \frac{1}{4}d\kappa  \cdot {\phi _ + }} \right|^2} + \Xi _ - ^2 + {\left| {\sigma \phi } \right|^2},
	\end{equation}
and therefore, we have the localization locus
	\begin{equation}
		\left\{ \begin{gathered}
  		F_A^ +  + h\left( {{\phi _ + }} \right) = 0 \hfill \\
  		{\slashed{D}_A}{\phi^A _ + } + \frac{1}{4}d\kappa  \cdot {\phi^A _ + } = 0 \hfill \\ 
		\end{gathered}  \right.,\;\;\;\;\;\;\;\;\left\{ \begin{gathered}
  		{\iota _R}{F_A} = 0 \hfill \\
  		{d_A}\sigma  = 0 \hfill \\
  		\Xi_ - ^{A = 1,2} = 0 \hfill \\
  		\sigma \left( {{\phi^A _ + }} \right) = 0 \hfill \\ 
		\end{gathered}  \right..
		\label{localization-locus}
	\end{equation}
Note that using the reality condition of $\phi_+^A$, the second equation on the left is equivalent to that of one component ${\phi _ + } \equiv \phi _ + ^{A = 1}$
	\begin{equation}
		{\slashed{D}_A}{\phi _ + } + \frac{1}{4}d\kappa  \cdot {\phi _ + } = 0
	\end{equation}
and similarly $\sigma \left( {{\phi _ + }} \right) = 0 \Leftrightarrow \sigma \left( {\phi _ + ^A} \right) = 0$. Therefore, in the following, we will just ignore the index $A$, and regard $\phi_+$ as in the fundamental representation of gauge group $G = U(N_c)$.

	\subsection{Deformed Coulomb Branch}

	The deformed Coulomb branch is the class of solutions to (\ref{localization-locus}) such that $\phi^A_+ = 0$. Then the equations reduces to
	\begin{equation}
		{d_A}\sigma  = 0,\;\;\;\;\;\;\;\;F_A^ +  - F_{{A_0}/2}^ +  = \frac{\zeta}{2} d{\kappa ^ + },\;\;\;\;\iota_R F_A = 0
	\end{equation}

	This is a deformed version of the contact-instanton equation introduced in \cite{Kallen:2012zr}. The undeformed version is later studied in \cite{Pan:2014aa, Harland:2011aa, Wolf:2012gz, Baraglia:2014aa}, in the context of $\kappa$ being a contact structure. So in principle, there could be a tower of instantonic solutions, very much like the deformed instantons in 4d.

	To be more concrete, we consider the case when $\kappa$ is a contact 1-form. Then $d\kappa^+ = d\kappa$, and one immediately has a most simple solution (assuming $\iota_R F_{A_0/2} = 0$)
	\begin{equation}
		A = \frac{\zeta}{2} \kappa + \frac{1}{2}A_0 
	\end{equation}
where $\sigma$ takes constant value in the Lie-algebra $\mathfrak{g}$. On top of these simple solutions, one may have a lot of instantonic solutions.

	When $(\kappa, R, g, \Phi)$ give rise to a Sasakian structure, the reference $A_0$ can be chosen to be the restriction on $K_M$ of the Chern connection on $K_{C\left( M \right)}$, where $C(M)$ is the Kahler cone of Sasakian manifold $M$. In such case, one can show that $d{A_0} \propto d\kappa $ and ${\iota _R}{F_{{A_0}/2}} = 0$.

	\subsection{5d Seiberg-Witten Equation}

	Let us consider other classes of solutions to (\ref{localization-locus}), with non-vanishing $\phi_+$. To be concrete in many statements, we will focus on the case where $(\kappa, R, g, \Phi)$ form a K-contact structure, or Sasakian structures to ensure concrete existence of solutions. This will allow us to rewrite the equations in a very geometric way that resembles the 4-dimensional Seiberg-Witten equation on symplectic manifolds. We will see that Sasakian structures serve as examples where Higgs vacua always exist, and other non-trivial solutions have nice behavior. We also discuss the case of general K-contact structures.
	
	\vspace{20pt}
	\underline{\emph{The algebraic equation}}

	When we look for non-vanishing solution of $\phi_+$, one of the non-trivial BPS equations is $\left( {\sigma  + m} \right)\left( \phi_+  \right) = 0$, where we have restored some masses for the hypermultplets by giving VEV to the scalars in the background vector multiplets that gauge the flavor symmetry. Let us consider gauge group $G = U(N_c)$ and $N_f$ hypermultiplets, then we need to solve a matrix equation $\left( {{\sigma ^a}_b + {m_i}^j} \right)\phi _j^b = 0$, where $a, b = 1, ..., N_c$ are gauge indices, while $i,j = 1, ..., N_f$ are flavor indices. After diagonalizing ${m^i}_j = \text{diag}(m_1, ..., m_{N_f})$, one observes that, assuming $N_c \le N_f$, any solution is determined by an ordered subset of integers $\left\{ {{n_1},...,{n_{{N_c}}}} \right\}$ of size $N_c$
	\begin{equation}
		{\sigma ^a}_b =  - {m_{{n_a}}}\delta_b^a,\;\;\;\;\phi _i^a\sim{\delta _{i,{n_a}}},\;\;\;\;\left\{ {{n_1},...,{n_{{N_c}}}} \right\} \subset \left\{ {1,...,{N_f}} \right\}.
		\label{algebraic}
	\end{equation}
Therefore $N_c$ among the $N_f$ of $\phi$'s are selected to have non-zero values. The remaining $N_f - N_c$ of $\phi$'s are fixed to be zero, and trivially satisfy all other BPS equations. These vanishing components do not have further non-trivial solutions which we will discuss shortly. The 1-loop determinants for the trivial components will be the same as that in the Coulomb branch, with the argument $\sigma$ replaced by solutions (\ref{algebraic}). 

	The selected $N_c$ ($< N_f$) non-zero components, on the other hand, requires extra care. First of all, given generic masses $\{m_{n_a} \ne m_{n_b}\; \text{if} \; a \ne b\}$, equation $d_A \sigma = 0$ implies $A$ is also completely diagonalized. Therefore, in such favorable situations, the gauge group $U(N_c)$ is completely broken to $U(1)^{N_c}$, which acts as phase rotations on the $N_c$ non-zero components of $\phi$. For each of these components, one only needs to consider a $U(1)$-gauge field, which we will assume from now on. These non-zero components will have to satisfy the remaining BPS equations individually, to which we will discuss the solutions shortly. To do so, we will first rewrite the remaining BPS equations in a more familiar form.

	\vspace{20pt}
	\underline{\emph{Rewriting the localization locus}}

	In the appendix [\ref{appendix-c}][\ref{appendix-d}], we review in detail $\operatorname{Spin}^\mathbb{C}$ spinors and corresponding Dirac operators on any 5-dimensional K-contact structures. We summarize here several most relevant aspects:
	\begin{itemize}
		\item The spinor bundle $S$ has a canonical Dirac operator $\slashed{\nabla}^\text{TW}$, induced from generalized Tanaka-Webster connection on $TM$ for any given K-contact structure\cite{Petit2005229}\cite{Nicolaescu:aa}\cite{Degirmenci:2013aa}. One can show that this Dirac operator can be written in terms of the Levi-Civita connection $\nabla^\text{LC}$:
		\begin{equation}
			{\slashed{\nabla} ^{{\text{TW}}}} = {\slashed{\nabla} ^{{\text{LC}}}} + \frac{1}{8}{\left( {d\kappa } \right)_{mn}}{\Gamma ^{mn}} \Rightarrow \left\{ \begin{gathered}
  		{P_ - }{\slashed{\nabla} ^{{\text{TW}}}}{\phi _ + } = {P_ - }{\slashed{\nabla} ^{{\text{LC}}}}{\phi _ + } \hfill \\
  		{P_ + }{\slashed{\nabla} ^{{\text{TW}}}}{\phi _ + } = {P_ + }{\slashed{\nabla} ^{{\text{LC}}}}{\phi _ + } + \frac{1}{4}d\kappa  \cdot {\phi _ + } \hfill \\
  		\;\;\;\;\;\;\;\;\;\;\;\;\;\;\;\;\; = - \left({\nabla^\text{LC} _R}{\phi _ + } + \frac{1}{4}d\kappa  \cdot {\phi _ + } \right ) \hfill \\ 
		\end{gathered}  \right.
		\end{equation}
		which are precisely the ones appearing in $Q\psi_\pm$ without the gauge field $A$.
		\item There exists a \emph{canonical} $\operatorname{Spin}^\mathbb{C}$-bundle ${W^0} = {T^{0, \bullet }}M_H^*$, with chiral decomposition
		\begin{equation}
			W_ + ^0 = {T^{0,0}}M_H^* \oplus {T^{0,2}}M_H^*,\;\;\;\;W_ - ^0 = {T^{0,1}}M_H^* 	
		\end{equation}
		and determinant line bundle $ K_M \equiv {T^{0,2}}M_H^*$. Any other $\operatorname{Spin}^\mathbb{C}$-bundle $W$ can be written as $W = {W^0} \otimes E$ for some $U(1)$-line bundle $E$. It is important to note that, when the manifold is spin, namely when the genuine spinor bundle exists, then $S$ and $W^0$ is related by $S \otimes K_M^{1/2} = W^0$ $ \Rightarrow S_+ = K_M^{ - 1/2} \otimes K_M^{1/2}$. Therefore $W$ can also be written as $W = S \otimes \mathcal{L}$ where $\mathcal{L} = K_M^ {1/2} \otimes E$.
		\item On $K_M$  there exists a \emph{canonical} $U(1)$ connection $A_0$, such that the Dirac operator (induced from $\nabla^\text{TW}$ on $TM$ and $A_0/2$ on $K_M^{1/2}$) on the canonical $\operatorname{Spin}^\mathbb{C}$-bundle $W^0$ satisfies the identity\footnote{It is the restriction onto $K_M^{-1}$ of the Chern connection defined on $T{C(M)}$, where $C(M)$ is the almost hermitian cone over the K-contact 5-manifold $M$; however, there are other choices (induced by $\nabla^\text{TW}$ discussed in \cite{Nicolaescu:aa}, for instance) of $A_0$ that leads to similar identification, with the only difference that $\mathcal{L}_R$ is replaced by ${\mathcal{L}_R} - i{a_0}\left( R \right)$ for some appropriate $U(1)$ gauge field $a_0$.}
		\begin{equation}
			D_{{A_0}/2}^{{\text{TW}}} = {\mathcal{L}_R} \oplus \sqrt 2 \left( {\bar \partial  + {{\bar \partial }^*}} \right):{\Omega ^{0,{\text{even}}}} \to {\Omega ^{0,{\text{even}}}} \oplus {\Omega ^{0,{\text{odd}}}}
		\end{equation}
	\end{itemize}

	Now we can include the gauge field $A$ onto the stage. As discussed above, we only consider $G = U(1)$ and $A$ is viewed as a $U(1)$-connection of certain line bundle $\mathcal{L}$. Therefore, $\phi_+$ should be really considered as a section of $W_+ \equiv S_+ \otimes \mathcal{L}$. We decompose $\mathcal{L} = K_M^{1/2} \otimes E$ so that $S \otimes \mathcal{L} = {W^0} \otimes E$, and we also decompose the gauge field $A$ according to
	\begin{equation}
		\begin{gathered}
  			{\phi _ + } \in W_0^ +  \otimes E = {S_ + } \otimes\;\; K_M^{1/2}\;\; \otimes \; E \hfill \\
  			\;\;\;\;\;\;\;\;\;\;\;\;\;\;\;\;\;\;\;\;\;\;\;\;\;\;\;\;\;\;\;\;\;\;\;\;\;A_0/2\;\; + \;\;a\;\; = \;\;A. \hfill \\ 
			\end{gathered}
	\end{equation}
Therefore, the Dirac operator $\slashed{D}_A^\text{TW}$ on $W_+ = W^+_0 \otimes E$ can be identified as 
	\begin{equation}
		{\slashed{D}_A} + \frac{1}{8}d{\kappa _{mn}}{\Gamma ^{mn}} = \slashed{D}_A^{{\text{TW}}} = \mathcal{L}_R^a \oplus \sqrt 2 \left( {{{\bar \partial }_a} + \bar \partial _a^*} \right):{W_ + } \to {W_ + } \oplus {W_ - }.
	\end{equation}
where $\mathcal{L}_R^a = {\mathcal{L}_R} - ia\left( R \right),\;\;{{\bar \partial }_a} = \bar \partial  - i{a^{0,1}}$ and so forth.

	With such identification in mind, one can rewrite the Dirac-like equation in (\ref{localization-locus})
	\begin{equation}
		{\slashed{D}_A}{\phi _ + } + \frac{1}{8}d{\kappa _{mn}}{\Gamma ^{mn}}{\phi _ + } = \slashed{D}_A^{{\text{TW}}}{\phi _ + } = 0 \Leftrightarrow \mathcal{L}_R^a{\phi _ + } = 0,\;\;\;\;\left( {{{\bar \partial }_a} + \bar \partial _a^*} \right){\phi _ + } = 0.
	\end{equation}
 In particular, we write ${\phi _ + } = \alpha  \oplus \beta  \in {\Omega ^{0,0}}\left( {{E}} \right) \oplus {\Omega ^{0,2}}\left( {{E}} \right)$, and (\ref{localization-locus}) can be written as
	\begin{equation}
		\left\{ \begin{gathered}
  		F_a^{d\kappa } = \frac{1}{2}\left( {\zeta  - {{\left| \alpha  \right|}^2} + {{\left| \beta  \right|}^2}} \right)d\kappa  \hfill \\
  		F_a^{0,2} = 2i\bar \alpha \beta  \hfill \\
  		{{\bar \partial }_a}\alpha  + \bar \partial _a^*\beta  = 0 \hfill \\ 
  		\mathcal{L}_R^a\alpha  = \mathcal{L}_R^a\beta  = 0 \hfill \\
		\end{gathered}  \right.,\;\;\;\;\;\;\left\{ \begin{gathered}
  		{\iota _R}{F_a} + {\iota _R}{F_{{A_0}/2}} = 0 \hfill \\
  		{d_A}\sigma  = 0 \hfill \\
 		{\Xi^{A = 1,2}_ - } = 0 \hfill \\
  		\sigma \left( \alpha  \right) = \sigma \left( \beta  \right) = 0 \hfill \\ 
		\end{gathered}  \right.
		\label{5d-SW}
	\end{equation}
where we have decompose $F_a^ +  = F_a^{d\kappa } + F_a^{2,0} + F_a^{0,2}$, and the bilinear map $\alpha(\phi)$ is written more concretely as (see appendix [\ref{appendix-a}, \ref{appendix-d}] for choice of basis and matrix representation of $\Gamma_{AB}$)
	\begin{equation}
		\alpha \left( \phi  \right) \equiv \frac{1}{2}\left( {{{\left| \alpha  \right|}^2} - {{\left| \beta  \right|}^2}} \right) d\kappa + 2i\left( {\alpha \bar \beta  - \bar \alpha \beta } \right),
		\label{alpha}
	\end{equation}
It is clear that the equations on the \emph{left} take a similar form of $\zeta$-perturbed Seiberg-Witten equations on a symplectic 4-manifold\cite{Taubes1996, Witten:1994cg,Taubes1997}, and therefore we will call them the \emph{5d Seiberg-Witten equations} in the following discussion.
	
	Let us pause to remark that, the operator $\slashed{\nabla} + 1/8d{\kappa _{mn}}{\Gamma ^{mn}}$ is discussed in the context of Sasaki-Einstein manifold, and similar results were obtained in \cite{Qiu:2013aa}. The unperturbed version of Seiberg-Witten-like equation on a contact metric manifold is also proposed in \cite{Degirmenci:2013aa}.

	\vspace{20pt}

	In the following we will focus on equations on the left in (\ref{5d-SW}). They are a novel type of equations that awaits more study. Let us try to make a first step to understanding the solutions. As discussed earlier, we consider the gauge group $G = U(1)$, and therefore $\sigma$ and $\zeta$ are just real constants.

	\underline{\emph{A Higgs vacuum}}

	First, we argue that the 5d Seiberg-Witten equations on Sasakian structures have one simple solution. 

	First of all, on any K-contact structure, $(\alpha, \beta) = (\sqrt{\zeta}, 0)$, together with $a = 0$, or equivalently $A = 1/2 A_0$, is obviously a solution to the 5d Seiberg-Witten equations.

	The remaining BPS equation is
	\begin{equation}
		{\iota _R}{F_{{A_0}/2}} = 0
		\label{horizontal}
	\end{equation}
If $A_0$ is chosen to be induced from 6d Chern connection, this may be not true on a general K-contact background; however, if the K-contact structure is Sasakian, then (\ref{horizontal}) indeed holds \cite{Degirmenci:2013aa}\cite{Petit2005229}. Therefore on a Sasakian structure, one always has at least one most simple solution, which we will call a \emph{Higgs vacuum}.

	\underline{\emph{Properties of general solutions}}

	Let us now focus on the 5d Seiberg-Witten equations on a K-contact structure (with emphasis on Sasakian structures).  First of all, the Dirac equations imply
	\begin{equation}
		\begin{gathered}
  		\;\;\;\;\;{{\bar \partial }_a}{{\bar \partial }_a}\alpha  + {{\bar \partial }_a}\bar \partial _a^*\beta  = 0 \Rightarrow  - iF_a^{0,2}\alpha  - N\left( {{\partial _a}\alpha } \right) + {{\bar \partial }_a}\bar \partial _a^*\beta  = 0 \hfill \\[0.5em]
   		\Rightarrow 2\int_M {{{\left| { \alpha } \right|}^2}{{\left| \beta  \right|}^2}}  - \int_M {\beta  \wedge {*_\mathbb{C}}N\left( {{\partial _a}\alpha } \right)}  + \int_M {{{\left| {\bar \partial _a^*\beta } \right|}^2}}  = 0. \hfill \\ 
		\end{gathered} 
	\end{equation}
where $N$ is the Nijenhuis tensor $N:{T^{1,0}}M_H^* \to {T^{0,2}}M_H^*$, which vanishes for any Sasakian structure. Therefore, when $(\kappa, R, g, \Phi)$ is Sasakian, one has
	\begin{equation}
		\bar \partial _a^*\beta  = {{\bar \partial }_a}\alpha  = \left| \alpha  \right|\left| \beta  \right| = 0.
	\end{equation}
Namely, either $\alpha$ or $\beta$ must vanish, and the two types of solutions are
	\begin{equation}
		\text{Sasakian:}\;\;\;\;\left\{ \begin{gathered}
  		\beta  = 0 \hfill \\
  		{{\bar \partial }_a}\alpha  = 0 \hfill \\ 
		\end{gathered}  \right.\;\;\;\;{\text{or}}\;\;\;\;\left\{ \begin{gathered}
  		\alpha  = 0 \hfill \\
  		\bar \partial _a^*\beta  = 0 \hfill \\ 
		\end{gathered}  \right..
		\label{sol-Sasakian}
	\end{equation}
However, unlike the case of 4-dimensional Kahler manifold, at the moment we do not have a topological characterization of the two types of solutions. Let us consider the curvature equation integrated over $M$
	\begin{equation}
		\int_M {F_a^{d\kappa } \wedge *d\kappa }  = \int_M {F_a^{d\kappa } \wedge \kappa  \wedge d\kappa }  = \frac{1}{2}\int_M {\left( {\zeta  - {{\left| \alpha  \right|}^2} + {{\left| \beta  \right|}^2}} \right)d\kappa  \wedge *d\kappa } .
	\end{equation}
In the case of a 4-dimensional Kahler manifold, the left hand side would be replaced by the intersection number ${c_1}\left( E \right) \cdot \left[ \omega  \right]$, a topological number independent on $\zeta$. Therefore, when $\zeta = 0$, the sign of ${c_1}\left( E \right) \cdot \left[ \omega  \right]$ will determine whether $\alpha$ or $\beta$ will survive; in particular, in the limit $\zeta \gg + 1$, only the solutions with $\beta = 0$ survive. On a 5-dimensional Sasakian manifold, however, the left hand side is not a topological number, and therefore at the moment we do not have a topological criteria to determine which of the (\ref{sol-Sasakian}) will survive.

	For non Sasakian K-contact structure, one needs to take the Nijenhuis tensor into account. Combining the Weitzenbock formula, Kahler identities and triangle inequalities, we obtain several estimates (where we rescaled $(\alpha, \beta)\to (\sqrt{\zeta}\alpha, \sqrt{\zeta}\beta)$, $z$ is some constant, and $\lambda > 1$ is a real constant)
	\begin{equation}
		\begin{gathered}
  		2\int_M {F_a^{d\kappa} \wedge *d\kappa }  \geqslant \left( {1 - \frac{{2z}}{\zeta }} \right)\int_M {{{\left| {d_a^J\alpha } \right|}^2}}  \hfill \\
  		\;\;\;\;\;\;\;\;\;\;\;\;\;\;\;\;\;\;\;\;\;\;\;\;\;\;\;\;\; + 2\zeta \int_M {{{\left( {1 - {{\left| \alpha  \right|}^2}} \right)}^2}}  + 2\zeta \int_M {{{\left| \alpha  \right|}^2}{{\left| \beta  \right|}^2}}  + 2\zeta \left( {1 - \frac{1}{\lambda }} \right)\int_M {{{\left| \beta  \right|}^2}}  \hfill \\ 
		\end{gathered} ,
	\end{equation}
and
	\begin{equation}
		\int_M \rho_{A_0} \left| \beta \right|^2 + \frac{1}{2}\int_M {{{\left| {\nabla_{A_0 + a}\beta } \right|}^2}}  + \zeta \int {{{\left| \beta  \right|}^4}}  + \frac{\zeta }{2}\int {{{\left| \beta  \right|}^2}}  < \frac{z}{\zeta }\int {{{\left| {d^J_a\alpha } \right|}^2}} ,
	\end{equation}
In the inequalities, $\nabla_{A_0 + a}$ is the connection on $K_M \otimes E$, $\rho_{A_0}$ is some function depending on $A_0$ but not on $\zeta$. Again, \emph{if} the integral on the left in the first estimate is bounded from above, or it scales at most of order ${\zeta ^{\epsilon  < 1}}$ ($\epsilon = 0$ in 4-dimension, since it is topological and independent on $\zeta$), then the above estimate tells us as $\zeta \to + \infty$, almost everywhere on $M$
	\begin{equation}
		\left| \beta  \right| \to 0,\;\;\left| \alpha  \right| \to 1,
	\end{equation}
and $\left| {d_J^a\alpha } \right|$ does not grow faster than $\zeta$. The second estimate then implies the overall derivative ${\nabla _{{A_0} + a}}\beta  \to 0$ faster than $\zeta^{\epsilon - 1}$, and therefore $\left| {\bar \partial _a^*\beta } \right| = \left| {{{\bar \partial }_a}\alpha } \right| \to 0$ as well.

	Therefore, let us make a bold conjecture that we have a similar situation as in 4-dimension. Namely for a general K-contact manifold, as $\zeta \to + \infty$, $\beta$ is highly suppressed, and we are left with $\alpha$ satisfying $\bar \partial_a \alpha = 0$, which approaches $\alpha \to 1$ rapidly once away from any zeros $\alpha^{-1}(0) \in M$. In the case of Sasakian manifold, the type of solutions with non-zero $\beta$ are less and less likely to survive when $\zeta \to +\infty$. With this conjecture in mind, we study the local behavior of 5d Seiberg-Witten equations with large positive $\zeta$ near any closed Reeb orbit.

	\subsection{The Local Model Near Closed Reeb Orbits \label{section-2.5}}

	On a generic contact manifold, the integral curve of the Reeb vector field may have uncontrollable behavior, as we mentioned early on. However, if the structure is K-contact, then the contact flow, viewed as a subgroup of the group $\operatorname{Isom}(M,g)$ of isometries, has a closure of $T^k \subset \operatorname{Isom}(M,g)$. 

	In other words, the integral curve of the Reeb vector field going through a point $p \in M$ forms a torus of dimension less than or equal to $k$. One can think of the curves as similar to irrational flows on a torus. The integer $k \le 3$ for a K-contact five-manifold, and is called the \emph{rank} of the structure. So, a rank-1 K-contact structure is a quasi-regular or regular contact structure, and $k \ge 2$ are all irregular.

	The isometric $T^k$-action highly degenerates at the closed Reeb orbits, namely $k - 1$ of the generators do nothing to the points on closed Reeb orbits. Therefore, at a small neighborhood $\mathcal{C} \times {\mathbb{C}^2}$ of a closed Reeb orbits $\mathcal{C}$, the $k - 1$ generators rotates the $\mathbb{C}^2$ (leaving $\mathcal{C}$ fixed), while the remaining 1 generator, corresponding to the Reeb field $R$, translates along $\mathcal{C}$.

	Bearing this picture in mind, one can write down an adapted coordinate $(\theta, z_1, z_2)$ on a small neighborhood $ \mathcal{C}\times\mathbb{C}^2$ of any closed orbit $\mathcal{C}$, such that $T^k = \{t_0, ..., t_{k-1}\}$ acts on it in an intuitive way. Such a coordinate system is characterized by the numbers $\left( {{\lambda _0};{\lambda _j},{m_{1j}},{m_{2j}}} \right)$, $j = 1,...,k-1$, where $\lambda_0, ..., \lambda_j$ are rationally independent positive real numbers, $m_{1j}$ and $m_{2j}$ are two lists of integers. In such a coordinate, the Reeb vector $R$ and contact 1-form $\kappa$ can be written as
	\begin{equation}
		\left\{ \begin{gathered}
  		R = {\lambda _0}\frac{\partial }{{\partial \theta }} + i \sum\limits_{i = 1,2} {\sum\limits_{j = 1}^{k - 1} {{\lambda _j}{m_{ij}}} \left( {{z_i}\frac{\partial }{{\partial {z_i}}} - {{\bar z}_i}\frac{\partial }{{\partial {{\bar z}_i}}}} \right)}  \hfill \\
  		\kappa  = \frac{1}{{{\lambda _0}}}\left( {1 - \sum\limits_{i = 1,2} {\sum\limits_{j = 1}^{k - 1} {{\lambda _j}{m_{ij}}} {{\left| {{z_i}} \right|}^2}} } \right)d\theta  + \frac{i}{2}\sum\limits_{i = 1,2} {{z_i}d{{\bar z}_i} - {{\bar z}_i}d{z_i}}  \hfill \\ 
		\end{gathered}  \right.
	\end{equation}
The isometric subgroup $T^k$ acts on the patch by
	\begin{equation}
		\left( {{t_0},{t_1},...,{t_{k - 1}}} \right) \cdot \left( {{e^{i\theta }},{z_1},{z_2}} \right) = \left( {{t_0}{e^{i\theta }},\prod\limits_{j = 1}^{k - 1} {t_j^{{m_{1j}}}{z_1}} ,\prod\limits_{j = 1}^{k - 1} {t_j^{{m_{2j}}}{z_2}} } \right)
	\end{equation}

	Let us pick a basis for horizontal 1-forms in region $\mathcal{C} \times \mathbb{C}^2$ 
	\begin{equation}
		e^5 = \kappa, \;\;\;\;{e^{{z_i}}} \equiv d{z_i} - i\frac{{{\Lambda _i}}}{{{\lambda _0}}}{z_i}d\theta ,\;\;\;\;{e^{{{\bar z}_i}}} \equiv d{\bar z_i} + i\frac{{{\Lambda _i}}}{{{\lambda _0}}}{\bar z_i}d\theta ,
	\end{equation}
where ${\Lambda _i} \equiv \sum\nolimits_{j = 1}^k {{\lambda _j}{m_{ij}}} $. It is straight-forward to show that ${\mathcal{L}_R}{e^{{z_i}}} = i{\Lambda _i}{e^{{z_i}}}$, ${\mathcal{L}_R}{e^{{{\bar z}_i}}} =  - i{\Lambda _i}{e^{{{\bar z}_i}}}$. One can also easily verify that $d\kappa = i e^{z_1}\wedge e^{\bar z_1} + i e^{z_2}\wedge e^{\bar z_2}$. This suggests that one can view $e^{z_i}$, $e^{\bar z_i}$ as spanning $T^{1,0}M^*$ and $T^{0,1}M^*$. Under such assumption, one can show $\forall\alpha \in \Omega^{0,0}$,
	\begin{equation}
		\left\{ \begin{gathered}
  		\partial \alpha  = \left( {{\partial _{{z_i}}}\alpha  + \frac{i}{2}{{\bar z}_i}{\mathcal{L}_R}\alpha } \right){e^{{z_i}}},\;\;\;\;\bar \partial \alpha  = \left( {{\partial _{{{\bar z}_i}}}\alpha  - \frac{i}{2}{z_i}{\mathcal{L}_R}\alpha } \right){e^{{{\bar z}_i}}} \hfill \\
  		\partial {e^{{z_i}}} = \frac{{{\Lambda _i}}}{2}{e^{{z_i}}} \wedge \left( {{{\bar z}_1}{e^{{z_1}}} + {{\bar z}_2}{e^{{z_2}}}} \right),\;\;\;\;\bar \partial {e^{{z_i}}} =  - \frac{{{\Lambda _i}}}{2}{e^{{z_i}}} \wedge \left( {{z_1}{e^{{{\bar z}_1}}} + {z_2}{e^{{{\bar z}_2}}}} \right) \hfill \\ 
		\end{gathered}  \right.
		\label{Dolbeault}
	\end{equation}

	\underline{\emph{Examples}}

	Let us look at the example of squashed $S^5 \subset \mathbb{C}^3$
	\begin{equation}
		S_\omega ^5 \equiv \left\{ {\left( {{z_1},{z_2},{z_3}} \right) \in {\mathbb{C}^3}|\sum\limits_{i = 1,2,3} {\omega _i^2{{\left| {{z_i}} \right|}^2}}  = 1} \right\}
	\end{equation}
One can define the Reeb vector field $R$ and contact 1-form $\kappa$ by restriction of
	\begin{equation}
		R \equiv i\sum\limits_{i = 1,2,3} {{\omega _i}\left( {{z_i}\frac{\partial }{{\partial {z_i}}} - {{\bar z}_i}\frac{\partial }{{\partial {{\bar z}_i}}}} \right)} ,\;\;\;\;\kappa  \equiv \frac{i}{2}\sum\limits_{i = 1,2,3} {\left( {{z_i}d{{\bar z}_i} - {{\bar z}_i}d{z_i}} \right)} 
	\end{equation}
Then it is easy to show that near the orbit $\mathcal{C}_3 \equiv \left\{ {\theta  \in \left[ {0,2\pi } \right]|\left( {0,0,{e^{i\theta }}\omega _3^{ - 1}} \right) \in S_\omega ^5} \right\}$, one can rewrite $R$ and approximate $\kappa$ in the new coordinate $\theta  = {\left( {2i} \right)^{ - 1}}\log \left( {{z_3}/{{\bar z}_3}} \right)$, ${w_i} \equiv \omega _3^{ - 1}\sqrt {{\omega _i}} {z_i}z_3^{ - 1}$.
	\begin{equation}
		\left\{ \begin{gathered}
  		R = {\omega _3}\frac{\partial }{{\partial \theta }} + i\sum\limits_{i = 1,2} {\left( {{\omega _i} - {\omega _3}} \right)\left( {{w_i}\frac{\partial }{{\partial {w_i}}} - {{\bar w}_i}\frac{\partial }{{\partial {{\bar w}_i}}}} \right)}  \hfill \\
  		\kappa  = \frac{1}{{{\omega _3}}}\left[ {1 - \sum\limits_{i = 1,2} {\left( {{\omega _i} - {\omega _3}} \right){{\left| {{w_i}} \right|}^2}} } \right]d\theta  + \frac{i}{2}\sum\limits_{i = 1,2} {{w_i}d{{\bar w}_i} - {{\bar w}_i}d{w_i}}  \hfill \\ 
		\end{gathered}  \right..
		\label{S5}
	\end{equation}
The natural $T^3$ action can be rearranged as
	\begin{equation}
		\left( {{e^{i\varphi }},{e^{i{\varphi _1}}},{e^{i{\varphi _2}}}} \right) \cdot \left( {{z_1},{z_2},{z_3}} \right) = \left( {{e^{i{\varphi _1}}}{e^{i\varphi }}{z_1},{e^{i{\varphi _2}}}{e^{i\varphi }}{z_2},{e^{i\varphi }}{z_3}} \right),
	\end{equation}
so that its action on the local coordinate is $\left( {{e^{i\varphi }}{e^{i\theta }},{e^{i{\varphi _1}}}{w_1},{e^{i{\varphi _2}}}{w_3}} \right)$, implying $m_{11} = m_{22} = 1$, and ${\lambda _{1,2}} = {\omega _{1,2}} - {\omega _3}$.

	Similar steps can be done on $Y^{pq} $ manifolds, which has K-contact rank $k = 2$. Let us recall how $Y^{pq}$ manifolds are defined \cite{Martelli:2004wu, Gauntlett:2004yd}. $Y^{pq}$ manifolds are Sasaki-Eintstein manifolds with topology $S^2 \times S^3$. They can be obtained by first looking at $S^3_{z_1, z_2} \times S^3_{z_3, z_4} \subset \mathbb{C}^4$ defined by equations
	\begin{equation}
		\left( {p + q} \right){\left| {{z_1}} \right|^2} + \left( {p - q} \right){\left| {{z_2}} \right|^2} = 1/2,\;p{\left| {{z_3}} \right|^2} + p{\left| {{z_4}} \right|^2} = 1/2
	\end{equation}
Then one can define a nowhere-vanishing $U(1)$-vector field $T$ which rotates the phases of $z_i$ according to the charges $[p+q, p-q, -p, -p]$. The $Y^{pq}$ manifolds is then the quotient $(S^3 \times S^3) / U(1)_T$. The Saaski-Einstein Reeb vector field is defined to be rotations of $z_i$ with irrational charges $[\omega_1, \omega_2, \omega_3, \omega_4]$
	\begin{equation}
		{\omega _1} = 0,\;\;\;\;{\omega _2} = \frac{1}{{\left( {p + q} \right)l}},\;\;\;\;{\omega _3} = {\omega _4} = \frac{3}{2} - \frac{1}{{2\left( {p + q} \right)l}}.
	\end{equation}
It is easy to show that near the closed Reeb orbit $\mathcal{C} \equiv \left\{ {\left( {{z_i}} \right) \in {Y^{pq}}|{z_2} = {z_4} = 0} \right\}$, one has
	\begin{equation}
		{\lambda _0} = p{\omega _1} + \left( {p + q} \right){\omega _3},\;\;\;\;{\lambda _1} = 3,\;\;\;\;{m_{11}} = 1, \;\;\;\;m_{21} = 0.
		\label{Ypq}
	\end{equation}

	\vspace{10pt}
	\underline{\emph{The 5d Seiberg-Witten equation near $\mathcal{C}$}}

	We study the equations near a closed orbit $\mathcal{C}$. Again, we rescale $(\alpha, \beta) \to (\sqrt{\zeta}\alpha, \sqrt{\zeta}\beta)$ for a better looking equation:
	\begin{equation}
		F_a^ +  = \frac{\zeta }{2}\left( {1 - {{\left| \alpha  \right|}^2} + {{\left| \beta  \right|}^2}} \right)d\kappa ,\;\;\;\;F_a^{0,2} = 2i\zeta\bar \alpha \beta ,\;\;\;\;\mathcal{L}_R^a\alpha  = \mathcal{L}_R^a\beta  = 0,\;\;\;\;{\bar \partial _a}\alpha  + \bar \partial _a^*\beta  = 0
		\label{SW-eq-flat}
	\end{equation}
Using (\ref{Dolbeault}) and its underlying assumption, the last equation in (\ref{SW-eq-flat}) can be reduced to usual equation on $\mathbb{C}^2$, since $\mathcal{L}_R^a \alpha = \mathcal{L}_R ^a \beta = 0$,
	\begin{equation}
		{{\bar \partial }_a}\alpha  + \bar \partial _a^*\beta  = 0\;\;{\text{on }}{\mathbb{C}^2}.
	\end{equation}
However, as we discussed early on, we conjecture that when $\zeta \to + \infty$, $\beta, \nabla \beta \to 0$ and therefore the differential equations of $\alpha$ and $\beta$ reduce to the holomorphic equation on $\mathbb{C}^2$
	\begin{equation}
		{{\bar \partial }_a}\alpha  = 0, \;\;\;\; \zeta \to + \infty.
	\end{equation}
In this sense, the zero set of large-$\zeta$ 5d Seiberg-Witten solutions corresponds to pseudo-holomorphic objects in K-contact manifold $M$. Namely near orbit $\mathcal{C}$, $\alpha^{-1}(0)$ takes the form of $\mathcal{C} \times \Sigma$ where $\Sigma$ is ``pseudo-holomorphically'' mapped into $M$. Of course this is just a naive description and far from rigorous; more careful treatment is needed.

	There are known smooth solutions to the 4-dimensional Seiberg-Witten equations, which are lifts of 2-dimensional vortex solutions; however, there are more solutions that we do not yet know how to describe. Nevertheless, let us assume that $\alpha$ has the usual asymptotic behavior $\alpha  \to {e^{i{n_0}\theta }}{e^{i{n_1}{\varphi _1} + i{n_2}{\varphi _2}}}$, where $n_0 \in \mathbb{Z} $, $n_{1,2} \in \mathbb{Z}_{\ge 0}$ is required by holomorphicity and smoothness at the origin\footnote{Not all modes above are possible. The precise range of these integers requires global analysis of the solution, which we will discuss in later examples.}: near the origin, $\alpha \sim {e^{in_0\theta }}z_1^{{n_1}}z_2^{{n_2}}$. Therefore,
	\begin{equation}
		\mathcal{L}_R^a \alpha = {\mathcal{L}_R}\alpha  - ia\left( R \right)\alpha  = 0 \Leftrightarrow {\lambda _0}{n_0} + {n_1}\sum\limits_{j = 1}^{k - 1} {{\lambda _j}{m_{1j}}}  + {n_2}\sum\limits_{j = 1}^{k - 1} {{\lambda _j}{m_{2j}}}  = a\left( R \right)
	\end{equation}
Note that the winding number $n_{0,1,2}$ should be bounded by $\zeta$, similar to the situation in \cite{Benini:2013yva}. We demonstrate this on a Sasakian structure in the limit $\zeta \gg 1$. We consider the integral
	\begin{equation}
		\int_M {{F_a^{d\kappa }} \wedge *d\kappa }  = \frac{\zeta }{2}\int {\left( {1 - {{\left| \alpha  \right|}^2}} \right)d\kappa  \wedge *d\kappa }  \leqslant \frac{\zeta }{2}\operatorname{Vol}\left( \kappa  \right),
	\end{equation}
where $\operatorname{Vol}\left( \kappa  \right) \equiv \int {d\kappa  \wedge *d\kappa } $. On the other hand, if $E$ is a trivial line bundle and thus $a$ can be viewed as a global 1-form, 
	\begin{equation}
		\begin{gathered}
  		\int_M {{F_a^{d\kappa }} \wedge *d\kappa }  = \int_M {da \wedge *d\kappa }  = \int_M {da \wedge \kappa  \wedge d\kappa }  = \int_M {a \wedge d\kappa  \wedge d\kappa }  \hfill \\
  		\;\;\;\;\;\;\;\;\;\;\;\;\;\;\;\;\;\;\;\;\;\; = \int {\left( {{\iota _R}a} \right)\kappa  \wedge d\kappa  \wedge d\kappa }  \hfill \\ 
		\end{gathered} 
	\end{equation}
Notice that if we assume the connections $a$ invariant under $\mathcal{L}_R$, then
	\begin{equation}
		{\iota _R}{F_a} = 0 \Rightarrow {\mathcal{L}_R}a = d{\iota _R}a = 0,
	\end{equation}
which leads to a bound on the winding numbers
	\begin{equation}
		{\lambda _0}{n_0} + {n_1}\sum\limits_{j = 1}^{k - 1} {{m_{1j}}{\lambda _j}}  + {n_2}\sum\limits_{j = 1}^{k - 1} {{m_{2j}}{\lambda _j}} = \iota_R a  \leqslant \frac{\zeta }{2}
		\label{bound}
	\end{equation}
Later we will see that this bound corresponds to poles in the perturbative Coulomb branch matrix model. More general situation needs more careful treatment, and we leave it for future study.

\section{Partition Function: Suppression and Pole Matching \label{section-3}}

	Suppose one obtains a BPS solution to the localization locus (\ref{localization-locus}), then the contribution to the partition function from this particular solution is the product
	\begin{equation}
		{e^{ - {S_{{\text{cl}}}}}}Z_{{\text{1-loop}}}^{{\text{vect}}}Z_{{\text{1-loop}}}^{{\text{hyper}}},
	\end{equation}
where $\exp \left[ { - {S_{{\text{cl}}}}} \right]$ is the exponentiated action evaluated on the BPS solution. The 1-loop determinants are
	\begin{equation}
		Z_{{\text{1-loop}}}^{{\text{vect}}}Z_{{\text{1-loop}}}^{{\text{hyper}}} = \frac{{{\text{sde}}{{\text{t}}_{{\text{vect}}}}\left( { - i{\mathcal{L}_R} + i\left( {\sigma  + i{\iota _R}{A_{{\text{cl}}}}} \right)} \right)}}{{{\text{sde}}{{\text{t}}_{{\text{Hyper}}}}\left( { - i\nabla _R^{{\text{TW}}} + i\left( {\sigma  + i{\iota _R}{A_{{\text{cl}}}}} \right)} \right)}}
	\end{equation}
where we have shifted $\sigma \to - i \sigma$, and $A_\text{cl}$ denotes the value of $A$ as a solution to (\ref{localization-locus}). Let us denote for a moment ${\mathcal{H}_A} \equiv \nabla _R^{{\text{TW}}} - i{A (R)}$, which we recall is part of the Dirac operator $\slashed{D}_A^\text{TW}$.

	In the Coulomb branch, where one does not include the deformation $Q V_\text{mixed}$, one encounters the BPS equations as a ``decoupled'' system of differential equations
	\begin{equation}
		\left\{ \begin{gathered}
  		F_A^ +  = 0,\;\;\;\;{d_A}\sigma  = 0,\;\;\;\;{\iota _R}{F_A} = 0 \hfill \\
  		{\slashed{D}_A}{\phi _ + } + \frac{1}{8}d{\kappa _{mn}}{\Gamma ^{mn}}{\phi _ + } = 0,\;\;\;\;\sigma \left( {{\phi _ + }} \right) = 0,\;\;\;\;F_ - ^{A = 1,2} = 0 \hfill \\ 
		\end{gathered}  \right.
	\end{equation}
In \cite{Qiu:2013aa}, it is shown that on a Sasaki-Einstein geometry (or other geometry with a large scalar curvature), a solution $A$ to the first line will imply the second line has only trivial solution $\phi_+ = 0$; namely the operator $\slashed{D}^\text{TW}_A$, and in particular $\mathcal{H}_A$ does not have \emph{zero} as one of its eigenvalues. Let $i\lambda_{\mathfrak{m}} \ne 0$ be an eigenvalue of $\mathcal{H}_A$ labeled by some quantum numbers $\mathfrak{m}$, with the corresponding eigenstate $\phi_{\mathfrak{m}}$. Then
	\begin{equation}
		{\mathcal{H}_A}{\phi _\mathfrak{m}} = i{\lambda _\mathfrak{m}}{\phi _\mathfrak{m}}
	\end{equation}
This is equivalent to the statement ${\mathcal{H}_{A + \Delta {A_\mathfrak{m}}}}{\phi _\mathfrak{m}} = 0$, where the $\Delta {A_\mathfrak{m}}\left( R \right) = {\lambda _\mathfrak{m}}$. Namely, there exists certain new gauge field $A + \Delta {A_m}$ with $\Delta A_\mathfrak{m}(R) = \lambda_\mathfrak{m}$, such that $\mathcal{H}_{A + \Delta A_\mathfrak{m}}$ has zero eigenvalue. Of course, $A + \Delta A$ cannot be a solution to the original Coulomb branch BPS equations, but it could be a solution to some deformed BPS equations. In our case, they are precisely the Higgs branch BPS equations, where the $Q V_\text{mixed}$ is taken into account. Therefore, solutions to the Higgs branch equations are expected to correspond to poles in the Coulomb branch matrix model, which are factors of the form ${\left( {i\sigma  - i{\lambda _\mathfrak{m}}} \right)^{ - 1}}$ coming from the hypermultiplet determinant. We will see this more precisely later in this section.

	\subsection{Suppression of the Deformed Coulomb Branch}

	In this subsection, we will review the supersymmetric actions for vector and hypermultiplet, and show that it is possible to achieve suppression of perturbative deformed Coulomb branch as $\zeta \to + \infty$ when certain bounds on the Chern-Simons level and the hypermultiplet mass are satisfied. This allows two things for theories containing hypermultiplets and appropriate Chern-Simons level,:

	1) one can take a large $\zeta$ limit, and only focus on the contributions from 5d Seiberg-Witten solutions to the partition function.

	2) One can take the Coulomb branch matrix model, close the integration contour of $\sigma$, and identify each pole of the integrand with a 5d Seiberg-Witten solution. Note that this is possible when the integrand is suppressed when $\zeta \to \infty$, and this requires the presence of hypermultiplets.

	3) For theories that do not satisfy the bounds, the above two statement are not valid in general. For instance, for pure super-Yang-Mills theory, one cannot close the contour and rewrite the matrix integral into sum of residues, and the deformed Coulomb branch will persist in large $\zeta$ limit.

	\vspace{20pt}
	\underline{\emph{The supersymmetric actions}}

	The Super-Yang-Mills and hypermultiplet action can be obtained by taking rigid limit of supergravity action. The bosonic parts read
	\begin{equation}
		\begin{gathered}
  		{\mathcal{L}_{{\text{YM}}}} = {\text{tr}}\left[ {F \wedge *F - \mathcal{A} \wedge F \wedge F - {d_A}\sigma  \wedge *{d_A}\sigma  - 1/2{D_{IJ}}{D^{IJ}}} \right. \hfill \\[0.5em]
  		\;\;\;\;\;\;\;\;\;\;\;\;\;\;\;\;\left. { - 4u{\sigma ^2} + \sigma {\mathcal{F}^{mn}}{F_{mn}} + 2\sigma \left( {{t^{IJ}}{D_{IJ}}} \right) + \sigma {F_{mn}}{\mathcal{P}^{mn}}} \right] \hfill \\ 
		\end{gathered} 
	\end{equation}
	\begin{equation}
  		{\mathcal{L}_{{\text{Hyper}}}} = {\epsilon ^{IJ}}{\Omega _{AB}}{\nabla _m}\phi _I^A{\nabla ^m}\phi _J^B  - {\epsilon ^{I'J'}}{\Omega _{AB}}\Xi_{I'}^A\Xi_{J'}^B + {\epsilon ^{IJ}}{\Omega _{AB}}\left( {\frac{\mathcal{R}}{4} + h - \frac{1}{4}{\mathcal{P}_{mn}}{\mathcal{P}^{mn}}} \right)\phi _I^A\phi _J^B
	\end{equation}
Note that we use the original field variables to write the action, and it is straight forward to use the invertible twisting to convert to new field variables.

	One can also add in $Q$-invariant Chern-Simons terms for the vector multiplet \cite{Kallen:2012zr}, and we have made the shift $\sigma \to i\sigma$ stated earlier
	\begin{equation}
		{\mathcal{L}_{{\text{SC}}{{\text{S}}_5}}} = {\mathcal{L}_{{\text{C}}{{\text{S}}_5}}}\left( {A - i\sigma \kappa } \right) - \frac{i k}{{8{\pi ^2}}}{\text{tr}}\left( {\Psi  \wedge \Psi  \wedge \kappa  \wedge {F_{A - i\sigma \kappa }}} \right),
	\end{equation} 
	\begin{equation}
		{\mathcal{L}_{{\text{SC}}{{\text{S}}_{3,2}}}} = {\mathcal{L}_{{\text{C}}{{\text{S}}_{3,2}}}}\left( {A - i\sigma \kappa } \right) - i {\text{tr}}\left( {d\kappa  \wedge \kappa  \wedge \Psi  \wedge \Psi } \right),
	\end{equation}
where the pure Chern-Simons terms are
	\begin{equation}
		\left\{ \begin{gathered}
  		{\mathcal{L}_{{\text{C}}{{\text{S}}_5}}}\left( A \right) = \frac{ik}{{24{\pi ^2}}}{\text{tr}}\left( {A \wedge dA \wedge dA + \frac{3}{2}A \wedge A \wedge A \wedge dA + \frac{3}{5}A \wedge A \wedge A \wedge A \wedge A} \right) \hfill \\
  		{\mathcal{L}_{{\text{C}}{{\text{S}}_{3,2}}}}\left( A \right) = i {\text{tr}}\left( {d\kappa  \wedge \left( {A \wedge dA + \frac{2}{3}A \wedge A \wedge A} \right)} \right) \hfill \\ 
		\end{gathered}  \right.
	\end{equation}
The 5d Chern-Simons level $k$ is an integer. As noted in \cite{Kallen:2012zr}, $\mathcal{L}_{\text{SCS}_{3,2}}$ is not invariant under rescaling of $\kappa$, while $\mathcal{L}_{\text{SCS}_{5}}$ is invariant.

	\vspace{20pt}
	\underline{\emph{The classical contributions}}

	The deformed Coulomb branch equations are
	\begin{equation}
		{d_A}\sigma  = 0,\;\;\;\;F_A^ +  - F_{{A_0}/2}^ +  = \frac{\zeta}{2} d{\kappa ^ + },\;\;\;\;{\iota _R}{F_A} = 0
	\end{equation}
On a Sasakian background, $\iota_R F_{A_0/2} = 0$, the perturbative solutions are
	\begin{equation}
		A = \frac{1}{2}{A_0} + \frac{\zeta}{2}\kappa ,\;\;\;\;\sigma = \text{constant} \in \mathfrak{u}(N_c)
		\label{deformed-Coulomb-sol}
	\end{equation}

	Evaluated on (\ref{deformed-Coulomb-sol}), the actions discussed above give the classical perturbative contribution to the partition function. We are interested the asymptotic behavior of these contributions as $\zeta \to +\infty$.

	1) The two Chern-Simons terms contribute up to factors of order $\exp{O(\zeta)}$
	\begin{equation}
		\exp \left( {i{S_{{\text{SC}}{{\text{S}}_5}}} + i\mu {S_{{\text{SC}}{{\text{S}}_{3,2}}}}} \right) \to \exp \left[ { - {\rm tr} \left( {\frac{k}{{24{\pi ^2}}}{{\left( {\sigma  + \frac{i}{2}\zeta } \right)}^3} + i\mu {{\left( {\sigma  + \frac{i}{2}\zeta } \right)}^2}} \right){\text{vol}}\left( \kappa  \right)} \right]
	\end{equation}
where we denote the contact volume $\operatorname{Vol} \left( \kappa  \right) = \int_M {\kappa  \wedge d{\kappa ^ + } \wedge d{\kappa ^ + }}  = \int_M {d{\kappa ^ + } \wedge *d{\kappa ^ + }}  $, and $\mu$ is a real coupling constant.

	2) There is no classical contribution from $\mathcal{L}_\text{Hyper}$ since all fields in the hypermultiplet vanish.

	3) Finally, there is classical contribution from $\mathcal{L}_\text{YM}$. To evaluate it, one needs to consider the field redefinition ${H_{mn}} = 2F_{mn}^ +  + \left( {2\sigma {t^{IJ}} + {D^{IJ}}} \right){\left( {{\Theta _{IJ}}} \right)_{mn}} - 4\mathcal{F}_{mn}^ + $, the equation of motion of $H$ and BPS equation to solve $D_{IJ}$ in terms of $\sigma$
	\begin{equation}
		{H_{mn}} = F_{mn}^ +  + h{\left( \phi  \right)_{mn}}, \;\;\;\;\;\;F_{mn}^ +  + h{\left( \phi  \right)_{mn}} = 0.
	\end{equation}
Using some Fierz-identities, the field redefinition implies
	\begin{equation}
		{D_{IJ}} = \left( {{h_{mn}} + 2\mathcal{F}_{mn}^ + } \right){\left( {{\Theta _{IJ}}} \right)^{mn}} - 2\sigma {t_{IJ}}.
		\label{D-sol}
	\end{equation}
With this one can evaluate the classical contribution of super-Yang-Mills action. In the simplest case with $\mathcal{F} = \mathcal{P} = 0$ (namely on a Sasaki-Einstein background), we have
	\begin{equation}
		\exp \left[ { - {S_{{\text{YM}}}}} \right] = \exp \left[ { - \frac{1}{2}{\rm tr}{{\left( {\sigma  + \frac{i}{2}\zeta } \right)}^2}\operatorname{Vol}\left( \kappa  \right) + ...} \right]
	\end{equation}
where $...$ denotes $O(\zeta)$ terms involving $F_{A_0/2}$. So we see there are competing $\zeta^2$-dependent terms in the norm of the classical contribution when $\zeta \to + \infty$\footnote{Although we are focusing our discussion on $\zeta$-dependent terms, the $\zeta$-independent terms including ${\rm tr}\sigma^2$ are still present in the matrix model integral as $\zeta \to \infty$ as a convergence factor when integrating $\sigma$.}
	\begin{equation}
		\left| {{e^{ - {S_{{\text{YM}}}} + i{S_{{\text{SC}}{{\text{S}}_5}}} + i\mu {S_{{\text{SC}}{{\text{S}}_{3,2}}}}}}} \right|\sim\exp \left[ {\frac{1}{8}{\rm tr}\left( {1 + \frac{k}{{4{\pi ^2}}}\sigma } \right)\operatorname{Vol}(\kappa){\zeta ^2} } \right]
	\end{equation}

	On more general background with non-vanishing $\mathcal{F}$ and $\mathcal{P}$, the classical contribution from $\exp\{-S_\text{YM}\}$ has the same leading behavior of $\zeta^2$ as above, although the precise value will depend on the geometric background. The 1-loop determinant will be more complicated products of triple-sine function, 

	\vspace{20pt}
	\underline{\emph{The perturbative 1-loop contributions}}

	The perturbative 1-loop determinant from Coulomb branch was studied in \cite{Kallen:2012fk, Imamura:2012aa, Qiu:2013aa}. It was shown that the 1-loop determinant can be expressed in terms of triple sine functions $S_3(z|\omega)$, or their particular products.

	The triple sine function $S_3(z|\omega)$ with $\omega = (\omega_1, \omega_2, \omega_3)$ is defined as the regularized infinite product
	\begin{equation}
		{S_3}\left( {z| \omega} \right) \equiv \prod\limits_{{n_1},{n_2},{n_3} = 0}^{ + \infty } {\left( {\sum\limits_{i = 1,2,3} {\left( {{n_i} + 1} \right){\omega _i}}  - z} \right)\left( {\sum\limits_{i = 1,2,3} {{n_i}{\omega _i}}  + z} \right)} 
	\end{equation}
or in terms of generalized $\Gamma$-function $\Gamma_3(z|\omega_1, \omega_2, \omega_3)$:
	\begin{equation}
		{S_3}\left( {z|\omega} \right) \equiv \frac{1}{{{\Gamma _3}\left( {z|{\omega _1},{\omega _2},{\omega _3}} \right)\Gamma_3 \left( {{\omega _1} + {\omega _2} + {\omega _3} - z|{\omega _1},{\omega _2},{\omega _3}} \right)}}
	\end{equation}
What is most important to us is the asymptotic behavior of the triple-sine function: when $\omega_i > 0$, we have when $z \to \infty$ ($B_{3,3}$ are multiple Bernoulli functions, see \cite{Narukawa2003,Jimbo:1996ss})
	\begin{equation}
		\begin{gathered}
  		\log {S_3}\left( {z|\omega } \right) \equiv  - \frac{1}{{3!}}{B_{3,3}}\left( z \right)\left( {\log z + C} \right) - \frac{1}{{3!}}{B_{3,3}}\left( {\left| \omega  \right| - z} \right)\left( {\log \left( {\left| \omega  \right| - z} \right) + C} \right) \hfill \\[0.5em]
  		\;\;\;\;\;\;\;\;\;\;\;\;\;\;\;\;\;\;\;\;\;\; - \gamma {\zeta _3}\left( {0,z} \right) - \gamma {\zeta _3}\left( {0,\left| \omega  \right| - z} \right) + O\left( {{z^{ - 1}}} \right) + O\left( {{{\left( {\left| \omega  \right| - z} \right)}^{ - 1}}} \right) \hfill \\ 
		\end{gathered} 
	\end{equation}
which implies
	\begin{equation}
		{S_3}\left( {z|\omega } \right) \to \left\{ \begin{gathered}
  		\exp \left[ { - \frac{{i\pi }}{{3!}}\frac{{{z^3}}}{{{\omega _1}{\omega _2}{\omega _3}}} + O\left( {{z^2}} \right)} \right],\;\;\;\;\operatorname{Im} z > 0 \hfill \\[0.5em]
  		\exp \left[ {\frac{{i\pi }}{{3!}}\frac{{{z^3}}}{{{\omega _1}{\omega _2}{\omega _3}}} + O\left( {{z^2}} \right)} \right],\;\;\;\;\;\;\;\operatorname{Im} z < 0 \hfill \\ 
		\end{gathered}  \right.
	\end{equation}

	The 1-loop determinant from perturbative Coulomb branch computed in literatures are products (over weights $\mu \in \mathfrak{R}$ to which the hypermultiplet belong) of triple sine functions, with argument of the form
	\begin{equation}
		z = i\left\langle {\mu ,\sigma } \right\rangle  + im + N\left( \omega  \right).
	\end{equation}
Here $N(\omega)$ is a real constant determined by equivariant parameters\footnote{For the individual triple sine function to converge, $N(\omega)$ is required to have imaginary part, but as discussed in \cite{Qiu:2013ab}, after all ingredients are multiplied together, one can take the real limit.}. For us, $\mathfrak{R}$ is the fundamental or anti-fundamental representation of $U(N_c)$ gauge group.

	If we consider the deformed Coulomb branch, then what we need is to compute the super-determinant of
	\begin{equation}
		i{Q^2} = \nabla _R^{{\text{TW}}} - iA\left( R \right) - \sigma  = \nabla _R^{{\text{TW}}} - \left( {\sigma  + \frac{i}{2}\zeta  + {\text{const}}} \right)
	\end{equation}
from hypermultiplet\footnote{1-loop determinant of vector multiplet is not affected by $\zeta$}, which effectively shifts $\sigma \to \sigma + i\zeta/2 + \text{const}$ in the Coulomb branch 1-loop determinant. In the limit of large $\zeta$, each $S_3$ factor of the 1-loop determinant of hypermultiplet tends to
	\begin{equation}
		\begin{gathered}
  		\left| {S_3\left( {z|\omega } \right)} \right|\xrightarrow{{\left\langle {\mu ,\sigma } \right\rangle  + m > 0,\left| \zeta  \right| \to \infty }}\left| {\exp \left[ { - \frac{{i\pi }}{{6{\omega _1}{\omega _2}{\omega _3}}}{{\left( {i\left\langle {\mu ,\sigma  + \frac{{i\zeta {1_{{N_c} \times {N_c}}}}}{2}} \right\rangle  + im + {\text{constant}}} \right)}^3}} \right]} \right| \hfill \\[0.5em]
  		\;\;\;\;\;\;\;\;\;\;\;\;\;\;\xrightarrow{{{\text{leading terms}}}}\exp \left[ { \frac{\pi }{{8{\omega _1}{\omega _2}{\omega _3}}}\left( {\left\langle {\mu ,\sigma } \right\rangle  + m} \right){\zeta ^2}} \right] \hfill \\ 
		\end{gathered}
	\end{equation}
Similarly,
	\begin{equation}
		\begin{gathered}
  		\left| {S_3\left( {z|\omega } \right)} \right|\xrightarrow{{\left\langle {\mu ,\sigma } \right\rangle  + m < 0,\left| \zeta  \right| \to \infty }}\left| {\exp \left[ { - \frac{{i\pi }}{{6{\omega _1}{\omega _2}{\omega _3}}}{{\left( {i\left\langle {\mu ,\sigma  + \frac{{i\zeta {1_{{N_c} \times {N_c}}}}}{2}} \right\rangle  + im + {\text{constant}}} \right)}^3}} \right]} \right| \hfill \\[0.5em]
  		\;\;\;\;\;\;\;\;\;\;\;\;\;\;\xrightarrow{{{\text{leading terms}}}}\exp \left[ { - \frac{\pi }{{8{\omega _1}{\omega _2}{\omega _3}}}\left( {\left\langle {\mu ,\sigma } \right\rangle  + m} \right){\zeta ^2}} \right] \hfill \\ 
		\end{gathered} 
	\end{equation}
Note that this asymptotic result is different from that in 3d. In 3d, there is an overall $\pm1$ factor in the exponent, corresponding to how the $\mathfrak{u}(1)$ parts act on the specific weight, while here such factor is squared to $1$. This reflects the symmetry in the matter content, where the fundamental and anti-fundamental (or $\mathfrak{R}$ and $\bar {\mathfrak{R}}$ in general) appear in a symmetric way in the hypermultiplet.
	
	As a simplest example, consider $N_f$ massless hypermultiplets on $S^5$ charged under gauge group $G = U(1)$. They contribute 1-loop determinant at large $\zeta$
	\begin{equation}
		\sim \exp \left[ { - \frac{\pi }{{8}} N_f \left| \sigma + m  \right|{\zeta ^2}} \right],
	\end{equation}
so the overall $\zeta^2$-terms in the norm of the matrix model integrand is
	\begin{equation}
		\exp \left[ {\frac{1}{8}\left( {1 + \frac{k}{{4{\pi ^2}}}\sigma } \right)4\pi^3{\zeta ^2} - \frac{\pi }{{8}}N_f\left| \sigma + m  \right|{\zeta ^2}} \right].
	\end{equation}
Therefore there is a window of suppression as $\zeta \to + \infty$
	\begin{equation}
		- {N_f} < k < {N_f},\;\;\;\;\frac{{4{\pi ^2}}}{{g_{{\text{YM}}}^2{N_f}}} \le \left| m \right|,
		\label{supression-bound}
	\end{equation}
where we have reinstated the $g_{\rm YM}$ which was omitted in front of the Yang-Mills action. In the above, the bound on $k$ comes from the competing $\sigma$ and $|\sigma|$ as one integrates $\sigma$ from $-\infty \to +\infty$, while the bound on $m$ comes from negating the positive $\zeta^2$-term from the Yang-Mills action. Within the suppression window, when performing the full matrix integral, because the integrand as a meromorphic function of $\sigma$ falls of exponentially fast far way from the real line, one can close the contour in the upper half plane, picking up residues from the poles; or alternatively, one can deform the integration contour from $\mathbb{R}$ to $\mathbb{R} + i \zeta$, and collecting a residue each time the contour passes a pole.

Similar result can be obtained for squashed $S^5$, where the volume ${\operatorname{Vol}}\left( \kappa  \right) \propto {\left( {{\omega _1}{\omega _2}{\omega _3}} \right)^{ - 1}}$, which only contributes an overall factor of the partition function as $\zeta \to + \infty$. On $Y^{pq}$ manifolds, one needs to replace the 1-loop determinant with generalized triple-sine functions, which are products of original triple-sine functions, and we expect one will have a similar suppression window where the Chern-Simons level and the hypermultiplet mass are constrained as $\zeta \to +\infty$.

One can generalize the above result to other gauge groups with $U(1)$ factors. For instance, consider on squashed $S^5$ the gauge group $G$ having $U(1)$-generators $h_a$. Define $\zeta  = {\zeta ^a}{h_a}$. Let the hypermultiplets belong to representations $\mathfrak{R}_{f = 1, ..., N_f}$, and $\mu$ will denote weights in $\mathfrak{R}_f$. The eigen-value of $h_a$ on $\mu$, namely the $U(1)$ charge, is denoted by ${q^f_a} \equiv \left\langle {\mu ,{h_a}} \right\rangle $. The large-$\zeta$ behavior of the exponent of the integrand is
	\begin{equation}
		\sim\frac{\pi }{{8{\omega _1}{\omega _2}{\omega _3}}}\sum\limits_{a,b} {{\zeta ^a}{\zeta ^b}\left[ { {4{\pi ^2}{\text{tr}}\left( {{h_a}{h_b}} \right) + k{\text{tr}}\left( {\sigma {h_a}{h_b}} \right)} - \sum\limits_{f = 1}^{{N_f}} {\sum\limits_{\mu  \in {\Re _f}} {q_a^fq_b^f\left| {\left\langle {\mu ,\sigma } \right\rangle  + {m_f}} \right|} } } \right]} 
	\end{equation}
The suppression can be achieved if the representations and the masses are such that the above expression tends to $\exp[- \infty]$ as $\zeta_a \to  \pm\infty$ (with some choice of sign). For instance, when $G = U(N_c)$, and $N_f$ hypermultiplets in the fundamental $\underline{N_c}$, the above reduces to
	\begin{equation}
		\sim\frac{{\pi {\zeta ^2}}}{{8{\omega _1}{\omega _2}{\omega _3}}}\left[ {4{\pi ^2}{N_c} + k{\text{tr}}\left( \sigma  \right) - \sum\limits_{f = 1}^{{N_f}} {\sum\limits_{\mu  \in {\underline{N_c}}} {\left| {\left\langle {\mu ,\sigma } \right\rangle  + {m_f}} \right|} } } \right],
	\end{equation}
and therefore suppression can be achieved if
	\begin{equation}
		\left| k \right| < {N_f},\;\;\;\;\sum\limits_{f = 1}^{{N_f}} {{m_f}}  > \frac{{4{\pi ^2}}}{{g_{YM}^2}}.
	\end{equation}

Finally, we remark that the bound above is a sufficient bound, obtained by only looking at the norm of the integrand. To fully understand when suppression can actually be achieved and whether or not the bound can be relaxed, more careful analyses are required. Also, the meaning of the mass bound is not clear to the authors at the moment, and we hope to get a better understanding in the future.

	\subsection{Matching The Poles And The Shift\label{sectoin-4.3}}

	Similar to 3-dimensional Higgs branch localization \cite{Benini:2013yva}, if one performs the integral of the Coulomb branch matrix model by closing the contour appropriately, one picks up residues from the enclosed poles. Before checking the matching between poles and 5d Seiberg-Witten equation, let us first understand the operator $\nabla _R^{{\text{TW}}} - i{\iota _R}A$ properly.

	\vspace{20pt}
	\underline{\emph{The operator $\nabla _{R,A}^{{\text{TW}}}$ and $\mathcal{L}_R$}}

	Let $\phi_+  = \xi  \otimes {\sigma _{E}}$ be a section of $S_+\otimes E$, where $E$ is equipped with $A$ as a $U(1)$ connection\footnote{Namely, ${\nabla _A}{\sigma _E} =  - iA{\sigma _E} \Rightarrow \nabla _A^{{\text{TW}}}\left( {\xi  \otimes {\sigma _E}} \right) = \left( {{\nabla ^{{\text{TW}}}} - iA} \right)\xi  \otimes {\sigma _E}$}. 

	Equivalently, noting that $S_+ = K_M^{-1/2} \otimes
	K_M^{1/2}$, one can choose an appropriate section $\hat \sigma $ of $ K_M^{1/2}$, and rewrite $\phi_+  = \left( {\xi  \otimes {\hat \sigma}} \right) \otimes ( {\hat \sigma^{ - 1} \otimes {\sigma _E}} )$, where we have factored out a piece $\xi  \otimes {\hat \sigma} \in \Gamma \left( {W_ + ^0} \right)$. $\hat \sigma$ then provides the explicit connection 1-form for the abstract canonical connection ``$A_0$'' on $K_M$:
	\begin{equation}
		{\nabla _{{A_0}/2}}{\hat \sigma} =  - i\frac{{{A_0}}}{2}{\hat \sigma},
	\end{equation}
and hence
	\begin{equation}
		\nabla _{R,A}^{{\text{TW}}}{\phi _ + } = {\mathcal{L}_R}\left( {\xi  \otimes {\hat \sigma}} \right) \otimes (\hat \sigma^{ - 1} \otimes {\sigma _E}) - i\left( {{\iota _R}a} \right){\phi _ + },
	\end{equation}
where we have used $\nabla _{R,{A_0}/2}^{{\text{TW}}} = {\mathcal{L}_R}$ on $W^0_+$, $a = A - {A_0}/2$ as a connection on $E \otimes K_M^{-1/2}$.

	In the case where $A = 0$, namely the perturbative Coulomb branch solution, one has ${\iota _R}a =  - {\iota _R}{A_0}/2$ and therefore the shift in eigenvalues of $\nabla^\text{TW}_R$ and $\mathcal{L}_R$
	\begin{equation}
		\Delta \left( {\nabla _R^{{\text{TW}}},{\mathcal{L}_R}} \right) = \frac{i}{2}{\iota _R}{A_0}.
	\end{equation}

	On the other hand, one of the BPS equation reads
	\begin{equation}
		\nabla _{R,A}^{{\text{TW}}}{\phi _ + } = 0 \Leftrightarrow {\mathcal{L}_R}\left( {\xi  \otimes {\hat \sigma}} \right) \otimes (\hat \sigma^{ - 1} \otimes {\sigma _E})  = i\left( {{\iota _R}a} \right){\phi _ + }
	\end{equation}
As a section of ${T^{0,0}}{M^*} \oplus {T^{0,2}}{M^*}$, $\xi  \otimes {\hat \sigma}$ contributes eigenvalues of $\mathcal{L}_R$ of the form
	\begin{equation}
		{\lambda _0}{n_0} + {n_1}\sum\limits_{j = 1}^{k - 1} {{\lambda _j}{m_{1j}}}  + {n_2}\sum\limits_{j = 1}^{k - 1} {{\lambda _j}{m_{2j}}}, \;\;\;\;n_{0} \in \mathbb{Z}, n_{1,2} \in \mathbb{Z}_{\ge 0}.
	\end{equation}
corresponding to modes with asymptotic behavior $\sim{e^{i{n_0}}}z_1^{{n_1}}z_2^{{n_2}}$ near each closed Reeb orbit. Now the remaining puzzle is to determine the value of $\iota_R A_0$.

	\vspace{20pt}
	\underline{\emph{Squashed $S^5$ and $\iota_R A_0$}}

	As an example, let us consider matching the poles of 1-loop determinant on squashed $S^5$ with the local solutions to the 5d Seiberg-Witten equation. We will focus on the orbit $\mathcal{C}_3$ discussed before, and recall the formula (\ref{S5}).

	Note that one can define local orthonormal vielbein $e^A$ by first defining an orthonormal frame at $\theta = 0$, then use $R$ to translate them to \emph{almost} the whole $\mathcal{C}_3$. In particular, one can define $e^A$ in such a way that it is \emph{adapted} to and \emph{invariant} under the K-contact structure, namely ${\mathcal{L}_R}{e^A} = 0$. However, translating $e^A$ back to $\theta = 2\pi$ will in general disagree with the starting value. To obtain a  vielbein well-defined on $\mathcal{C}_3$, one can rotate the original $e^A$ along the way. For instance, in terms of the complex basis
	\begin{equation}
		{e^{{z_i}}} \to \exp \left( {i\frac{{{\omega _i} - {\omega _3}}}{{{\omega _3}}}\theta } \right){e^{{z_i}}},\;\;{e^{{{\bar z}_i}}} \to \exp \left( { - i\frac{{{\omega _i} - {\omega _3}}}{{{\omega _3}}}\theta } \right){e^{{{\bar z}_i}}}
	\end{equation}
Then we have
	\begin{equation}
		{\mathcal{L}_R}{e^{{{\bar z}_i}}} =  - i\left( {{\omega _i} - {\omega _3}} \right){e^{{{\bar z}_i}}} \Leftrightarrow \left\{ \begin{gathered}
  		{\mathcal{L}_R}{e^{2i - 1}} =  - \left( {{\omega _i} - {\omega _3}} \right){e^{2i}} \hfill \\
  		{\mathcal{L}_R}{e^{2i}} = \left( {{\omega _i} - {\omega _3}} \right){e^{2i - 1}} \hfill \\ 
		\end{gathered}  \right.
	\end{equation}
In this basis, one can compute the derivative along $R$
	\begin{equation}
		\nabla _R^{{\text{LC}}} \psi = {R^m}{\partial _m}\psi + \frac{1}{2}\sum\limits_{i = 1,2} {\left( {{\omega _i} - {\omega _2}} \right){\Gamma ^{2i - 1}}{\Gamma ^{2i}}}\psi  - \frac{1}{4}d\kappa  \cdot \psi
	\end{equation}
Let $\psi_+ = (a, b)^T \in S_+$. Using the explicit representation (\ref{gamma-rep}) the derivative $\nabla^\text{LC}_R$ reduces to
	\begin{equation}
		\nabla _R^{{\text{LC}}}{\psi _ + } = {R^m}{\partial _m}{\psi _ + } + \frac{1}{2}\sum\limits_{i = 1,2} {\left( {{\omega _i} - {\omega _3}} \right)} i{\sigma _3}{\psi _ + } - i{\sigma _3}{\psi _ + },
	\end{equation}
where we used ${\Gamma ^{12}} \psi_+ = {\Gamma ^{34}} \psi_+= i{\sigma _3}\psi_+$ and $d\kappa  \cdot {\psi _ + } = 4i \sigma_3 \psi_+$. 

	When $\omega_{1,2,3} = 1$, one can define Killing spinor by
	\begin{equation}
		\nabla _m^{{\text{LC}}}\xi  =  - \frac{i}{2}{\Gamma _m}\xi .
	\end{equation}
Suppose $\xi_{-1/2} \in K_M^{-1/2}$ is a solution to the above Killing spinor equation, then using the above local expression of $\nabla^\text{LC}$, one can show that $\xi $ behaves like $\sim\exp \left( {\frac{{3i}}{2}\theta } \right)$ along $\mathcal{C}_3$. Finally, if we require $\hat \sigma$ to satisfy
	\begin{equation}
		{\xi _{ - 1/2}} \otimes {\hat \sigma} = \text{Const}\in \Gamma(T^{0,0}M^*),
	\end{equation}
one deduces that along $\mathcal{C}_3$
	\begin{equation}
		{\nabla _{R,{A_0}/2}}\hat \sigma  =  - \frac{{3i}}{2}\hat \sigma  =  - \frac{i}{2}\left( {{\iota _R}{A_0}} \right)\hat \sigma,
	\end{equation}
namely, along $\mathcal{C}_3$, $\hat \sigma$ has periodic behavior $\exp( - \frac{3i}{2}\theta)$ to cancel that of $\xi_{-1/2}$. This implies the shift
	\begin{equation}
		\Delta \left( {\nabla _R^{{\text{TW}}},{\mathcal{L}_R}} \right) = \frac{i}{2}{\iota _R}{A_0} = \frac{3i}{2}.
	\end{equation}

	On a general squashed $S^5_\omega$, we continue to choose $\hat \sigma$ such that it has $\exp\left( - \frac{3i}{2}\theta\right)$ periodic behavior along all three closed Reeb orbits. Then near any of three orbits, we recover the shift of eigenvalues as in \cite{Qiu:2013ab}\cite{Imamura:2012aa}
	\begin{equation}
		\Delta \left( {\nabla _R^{{\text{TW}}},{\mathcal{L}_R}} \right) = \frac{i}{2}\iota_R A_0 = \frac{{i\left( {{\omega _1} + {\omega _2} + {\omega _3}} \right)}}{2}
	\end{equation}

	Finally, the bound (\ref{bound}) on the winding numbers can now be written as
	\begin{equation}
		\sum\limits_{i = 1,2,3} {\left( {{n_i} + \frac{1}{2}} \right){\omega _i}}  \leqslant \frac{\zeta }{2} + \frac{{{\iota _R}{A_0}}}{2}.
		\label{bound-2}
	\end{equation}
where we defined $n_3 = n_0 - n_1 - n_2$, which is non-negative if one consider all three closed Reeb orbits $\mathcal{C}_{1,2,3}$. Recall that the 1-loop determinant in deformed Coulomb branch is obtained by a shift in that of Coulomb branch
	\begin{equation}
		\sigma  \to \sigma  + i\left( {\frac{\zeta }{2} + \frac{{{\iota _R}{A_0}}}{2}} \right) \Leftrightarrow \operatorname{Im} \sigma  = \frac{\zeta }{2} + \frac{{{\iota _R}{A_0}}}{2}.
	\end{equation}
Combining with the (\ref{bound-2}), bound saturation then means
	\begin{equation}
		\operatorname{Im} \sigma  = \sum\limits_{i = 1,2,3} {\left( {{n_i} + \frac{1}{2}} \right){\omega _i}} , \;\;\;\;n_i \ge 0,
	\end{equation}

	\vspace{10pt}
	\underline{\emph{Poles of the $S^5_\omega$ perturbative 1-loop determinant}}
	
	Recall that the perturbative 1-loop determinant of a hypermultiplet coupled to a $U(1)$ vector multiplet on $S^5_\omega$ is
	\begin{equation}
		Z_{{\text{1-loop}}}^{{\text{Hyper}}}\left( {S_\omega ^5} \right) = \left[{S_3}{\left( {i\sigma  + im + \frac{{{\omega _1} + {\omega _2} + {\omega _3}}}{2}|\omega } \right)}\right]^{ - 1}
	\end{equation}
The poles are the zeros of the infinite products
	\begin{equation}
		\prod\limits_{n \geqslant 0} {\left( {\sum\limits_{i = 1,2,3} {\left( {{n_i} + \frac{1}{2}} \right){\omega _i}}  - i\left( {\sigma  + m} \right)} \right)} \prod\limits_{n \geqslant 0} {\left( {\sum\limits_{i = 1,2,3} {\left( {{n_i} + \frac{1}{2}} \right){\omega _i}}  + i\left( {\sigma  + m} \right)} \right)} ,
	\end{equation}
where we have reinstated the mass induced from a background $U(1)$ vector multiplet. All the possible poles are
	\begin{equation}
		- m \pm i\sum\limits_{i = 1,2,3} {\left( {{n_i} + \frac{1}{2}} \right){\omega _i}}  = \sigma  \Leftrightarrow \operatorname{Re} \sigma  =  - m,\;\;\operatorname{Im} \sigma  =  \pm \sum\limits_{i = 1,2,3} {\left( {{n_i} + \frac{1}{2}} \right){\omega _i}} ,
	\end{equation}

	The first equation above is just the equation $\left( {\sigma  + m} \right)\phi  = 0$ in the Higgs branch, and the second is just the bound we obtained above, if one takes the poles with $+$ sign. These are the poles that will be picked up when one close the contour in the upper half plane of the $\sigma$-plane. Note that this is allowed thanks to the suppression of deformed Coulomb branch as $\zeta \sim \operatorname{Im} \sigma  \to  + \infty $.

	\vspace{20pt}
	\underline{\emph{The case of $Y^{pq}$ manifolds}}

	Recall (\ref{Ypq}) that near the orbit ${z_2} = {z_4} = 0$, the Sasaki-Einstin Reeb vector field can be written as
	\begin{equation}
		\begin{gathered}
  		R = \left[ {p{\omega _1} + \left( {p + q} \right){\omega _3}} \right]\frac{\partial }{{\partial \theta }} \hfill \\
  		\;\;\;\;\;\;\;\; + i\left( {{\omega _2} + {\omega _1} + 2{\omega _3}} \right)\left( {{u_1}\frac{\partial }{{\partial {u_1}}} - {{\bar u}_1}\frac{\partial }{{\partial {{\bar u}_1}}}} \right) + i\left( {{\omega _4} - {\omega _3}} \right)\left( {{u_2}\frac{\partial }{{\partial {u_2}}} - {{\bar u}_2}\frac{\partial }{{\partial {{\bar u}_2}}}} \right) \hfill \\ 
		\end{gathered} 
	\end{equation}
where
	\begin{equation}
		\displaystyle {\omega _1} = 0,\;\;{\omega _2} = \frac{1}{{\left( {p + q} \right)l}},\;\;\;\;{\omega _3} = {\omega _4} = \frac{1}{2}\left( {3 - \frac{1}{{\left( {p + q} \right)l}}} \right).
	\end{equation}

	One can then read off again $\iota_R A_0 = 3$ by choosing the section $\hat \sigma$ with the same criteria as $S^5$, and the bound on local winding number is also determined
	\begin{equation}
		{n_0}\left( {\frac{3}{2}\left( {p + q} \right) - \frac{1}{{2l}}} \right) + 3{n_1} + \frac{3}{2} \le \frac{\zeta}{2} + \frac{1}{2}{\iota _R}{A_0}, \;\;\;\;n_0 \in \mathbb{Z}, n_1 \in \mathbb{Z}_{\ge 0}.
	\end{equation}
After redefinition ${n_{{e_1}}} \equiv {n_1} + {n_0}p$, $n_\alpha \equiv n_0$, the bound saturation corresponds to the poles\footnote{The involved generalized triple sine function is \cite{Qiu:2013ab}
	\begin{equation}
		\prod\limits_{\Lambda _n^ + } {\left[ {\sum\limits_{i = 1}^4 {\left( {{n_{{e_i}}} + \frac{1}{2}} \right){\omega _i}}  + i\left( {\sigma  + m} \right)} \right]} \prod\limits_{\Lambda _n^ - } {\left[ {\sum\limits_{i = 1}^4 {\left( {{n_{{e_i}}} + \frac{1}{2}} \right){\omega _i}}  + i\left( {\sigma  + m} \right)} \right]} ,
	\end{equation}
where $\Lambda^\pm_n$ denotes restrictions on $n_{e_i}$
	\begin{equation}
		\left\{ \begin{gathered}
  		{n_{{e_1}}} + {n_{{e_2}}} - {n_{{e_3}}} - {n_{{e_4}}} = {n_\alpha }q \hfill \\
  		{n_{{e_1}}} - {n_{{e_2}}} =  - {n_\alpha }p \hfill \\ 
		\end{gathered}  \right.,\;\;\;\;\left\{ \begin{gathered}
  		{n_{{e_i}}} \geqslant 0,\;\;\;\;n \in \Lambda _n^ +  \hfill \\
  		{n_{{e_i}}} < 0,\;\;\;\;n \in \Lambda _n^ -  \hfill \\ 
		\end{gathered}  \right.
	\end{equation}}
	\begin{equation}
		\operatorname{Im} \sigma = 3{n_{{e_1}}} + {n_\alpha }\left( {\frac{3}{2}\left( {q - p} \right) - \frac{1}{{2l}}} \right) + \frac{3}{2} .
	\end{equation}
We remark that the redefinition seems to implies $n_{e_1} \in \mathbb{Z}$, but global analysis, namely, the equation (71) in \cite{Qiu:2013ab} implies $n_{e_1} + n_\alpha p = n_{e_2} \ge 0$ for the poles in the upper-half $\sigma$ plane.

	\section{Summary}

	In this work, we apply the idea of Higgs branch localization to supersymmetric theories of $\mathcal{N} = 1$ vector and hypermultiplet on general K-contact background. We show that with this generality the localization locus are described by perturbed contact instanton equations in the deformed Coulomb branch, and 5d Seiberg-Witten equations in the Higgs branch. Neither of these two types of equations is well understood. We focused on the latter, and some study basic properties of its solutions, including their local behavior near closed Reeb orbits, which is shown to reduce to 4-dimension Seiberg-Witten equations. This seems to implies that these BPS solutions corresponds to ``pseudo-holomorphic'' objects in K-contact manifolds, if the 4-dimensional story can some how be lifted. Finally, we study the suppression of deformed Coulomb branch as the parameter $\zeta \to + \infty$, and manage to match the poles of perturbative Coulomb branch matrix model with the bound on local winding numbers.

	From this point on, it is straight-forward to use the factorization property of perturbative partition function on $S^5$ and $Y^{pq}$ manifolds to perform the contour integral of $\sigma$. The result should produce classical and 1-loop contributions of each local Seiberg-Witten solutions, in a form of products of contributions from each closed Reeb orbit.


	Another question that we did not address is that whether the partition function is invariants of certain geometric structure. In \cite{Imamura:2014aa}, it is shown that the generalized Killing spinor equation (\ref{Killing-eq-0}) has huge degrees of \emph{local} freedom, including the background metric $g$, $\kappa$ and $R$, which are reflected as $Q$-exact deformations in the partition function. Therefore it would be interesting to explore the geometric or topological meaning of $\mathcal{N} = 1$ partition functions and expectation values of BPS operators. We believe that one needs to look closely the constraint (\ref{dilatino-eq}) and understand its geometric meaning. Also, one can further study the 5d Seiberg-Witten equations (\ref{5d-SW}). For instance, it would be interesting to understand its moduli spaces, which we did not take into account when matching the poles. But it is likely that on generic K-contact structures, the moduli spaces are zero-dimensional, considering the matching of perturbative poles and local solutions. Another interesting question is whether the solutions to (\ref{5d-SW}) correspond to certain ``pseudo-holomorphic'' objects, similar to the 4-dimensional story. If so, the partition functions will have more explicit geometrical meaning in terms of a ``counting'' of these objects.

	Finally we have the issue of $A_0$. In several discussions, including obtaining the bound on winding number, we relied on the assumption that the K-contact structure is Sasakian, in order to have a simplification $\iota_R F_{A_0/2} = 0$. It is not clear if this can always be achieved on general K-contact structures, or if there are other wiser choice of $A_0$ with the horizontal property, while simultaneously enables the identification $\slashed{D}_{{A_0}/2}^{{\text{TW}}} \leftrightarrow {\mathcal{L}_R} + \left( {\bar \partial  + {{\bar \partial }^*}} \right)$.

	\vspace{20pt}
	\textbf{Acknowledgment}

	We thank Francesco Benini and Wolfger Peelaers for explaining their work in great detail. We thank Francesco Benini, Dario Martelli, Wolfger Peelaers, Martin Ro\v{c}ek and Maxim Zabzine for reading the manuscript and their helpful comments. We thank Sean Fitzpatrick for discussions on related mathematics. The author would also like thank NSF grant no. PHY-1316617 for partial support.
	
\begin{appendices}
	\section{Spinors and Gamma Matrices\label{appendix-a}}

	In this appendix we review our convention on spinors and Gamma matrices, as well as useful formula.

	\underline{\emph{Spinors and Gamma matrices}}

	First let us consider a 5-dimensional spin manifold $M$. The rank of spin bundle $S$ is ${\operatorname {rank}}_\mathbb{C}S = {2^{\left[ {5/2} \right]}} = 4$. The metric on $TM$ induces a Clifford multiplication, expressed by Gamma matrices $\Gamma_m$, such that $\left\{ {{\Gamma _m},{\Gamma _n}} \right\} = 2{g_{mn}}$. The charge conjugatoin matrix $C = C_+$ satisfies
	\begin{equation}
		C{\Gamma ^m} = {\left( {{\Gamma ^m}} \right)^T}C.
	\end{equation}
We use lower case Greek letters $\alpha, \beta, ...$ to denote spinor indices, and overline $\bar z$ to denote usual complex conjugation of any complex number $z$. The complex conjugate of a spinor is defined as ${\bar \xi ^\alpha } = \overline {{\xi ^\alpha }} $.

	We define
	\begin{equation}
		{\Gamma _{mn}} \equiv \frac{1}{2}\left( {{\Gamma _m}{\Gamma _n} - {\Gamma _n}{\Gamma _m}} \right)
	\end{equation}
and similarly for ${\Gamma _{mnk}}$, ${\Gamma _{mnkl}}$. These products of Gamma matrices satisfy
	\begin{equation}
		{\Gamma _{mnk}} =  - \frac{{\sqrt g }}{2}{\epsilon _{mnkpq}}{\Gamma ^{pq}},\;\;\;\;{\Gamma _{mnkl}} = \sqrt g {\epsilon _{mnklp}}{\Gamma ^p}
	\end{equation}

	One can define a chiral and anti-chiral decomposition using any unit-normed vector field. In our case, we use the Reeb vector field $R$ and define a chiral operator $\Gamma_C \equiv - R^m \Gamma_m$, and decompose $S = S_+ \oplus S_-$.

	An explicit representation of Gamma matrices we will use is
	\begin{equation}
		\begin{gathered}
  		{\Gamma ^1} = \left( {\begin{array}{*{20}{c}}
  		0&{ - {i\sigma ^3}} \\ 
  		{{i\sigma ^3}}&0 
		\end{array}} \right),\;\;{\Gamma ^2} = \left( {\begin{array}{*{20}{c}}
  		0&{ - I} \\ 
  		{ - I}&0 
		\end{array}} \right), \hfill \\
  		{\Gamma ^3} = \left( {\begin{array}{*{20}{c}}
  		0&{ - {i\sigma ^1}} \\ 
  		{{i\sigma ^1}}&0 
		\end{array}} \right),\;\;{\Gamma ^4} = \left( {\begin{array}{*{20}{c}}
  		0&{ - {i\sigma ^2}} \\ 
  		{{i\sigma ^2}}&0 
		\end{array}} \right), \hfill \\
		\end{gathered} \;\;\;\;{\Gamma ^5} = \left( {\begin{array}{*{20}{c}}
  		{ - I}&0 \\ 
  		0&{ + I} 
		\end{array}} \right)
		\label{gamma-rep}
	\end{equation}

	\underline{\emph{Symplectic-Majorana spinors}}

	As opposed to that in 4-dimension, one cannot impose simple Majorana condition on a 5d spinor $\xi$. But one can define a {\emph {symplectic-Majorana spinor}}, as a pair of spinors $\xi_I, I = 1,2$, such that
	\begin{equation}
		\overline {\xi _I^\alpha }  = {C_{\alpha \beta }}{\epsilon ^{IJ}}\xi _J^\beta.
	\end{equation}
Note that given any usual spinor $\xi$, one can upgrade it to the symplectic-Majorana version by setting ${\xi _{I = 1}} = \xi ,\;{\xi _{I = 2}} =  {C^{ - 1}}\bar \xi $.

	Using $C$, one can define a $\mathbb{C}$-valued anti-symmetric product of any two arbitrary spinors $\xi$ and $\chi$
	\begin{equation}
		\left( {\xi \chi } \right) \equiv \sum\limits_{\alpha ,\beta  = 1,2} {{\xi ^\alpha }{C_{\alpha \beta }}{\chi ^\beta }} \in \mathbb{C}.
	\end{equation}
The product satisfies (here we consider Grassmann even spinors)
	\begin{equation}
		\left( {\xi \chi } \right) =  - \left( {\chi \xi } \right),\;\;\left( {\xi {\Gamma _m}\chi } \right) =  - \left( {\chi {\Gamma _m}\xi } \right),\;\;\left( {\xi {\Gamma _{mn}}\chi } \right) = \left( {\chi {\Gamma _{mn}}\xi } \right)
	\end{equation}

	One can also define an $\mathbb{R}$-valued symmetric inner product on $S$. Let $\xi$ and $\chi$ be any two spinors, and we upgrade them to symplectic-Majorana spinor $\xi_I$ and $\chi_I$. Then the inner product $\left \langle, \right \rangle$ is defined as
	\begin{equation}
		\left\langle {\xi ,\chi } \right\rangle  \equiv  {\epsilon ^{IJ}}\left( {{\xi _I}{\chi _J}} \right) = \sum\limits_\alpha  {\xi _1^\alpha \overline {\chi _1^\alpha }  + \overline {\xi _1^\alpha } \chi _1^\alpha }  = \sum\limits_\alpha  {{\xi ^\alpha }\overline {{\chi ^\alpha }}  + \overline {{\xi ^\alpha }} {\chi ^\alpha }}  \in \mathbb{R}
	\end{equation}
In particular, if $\xi \ne 0$ then $\left\langle {\xi ,\xi } \right\rangle  = 2\sum\limits_\alpha  {\overline {{\xi ^\alpha }} {\xi ^\alpha }}  > 0$.

	\underline{\emph{Fierz identities}}

	For arbitrary Grassmann even spinors $\xi_{1,2,3}$, we have the basic Fierz identity
	\begin{equation}
		{\xi _1}\left( {{\xi _2}{\xi _3}} \right) = \frac{1}{4}{\xi _3}\left( {{\xi _2}{\xi _1}} \right) + \frac{1}{4}{\Gamma ^m}{\xi _3}\left( {{\xi _2}{\Gamma _m}{\xi _1}} \right) - \frac{1}{8}{\Gamma ^{mn}}{\xi _3}\left( {{\xi _2}{\Gamma _{mn}}{\xi _1}} \right)
	\end{equation}
It follows immediately two useful formula
	\begin{equation}
		\left\{ \begin{gathered}
  		{\xi _1}\left( {{\xi _2}{\xi _3}} \right) + {\xi _2}\left( {{\xi _1}{\xi _3}} \right) =  - \frac{1}{4}{\Gamma ^{mn}}{\xi _3}\left( {{\xi _2}{\Gamma _{mn}}{\xi _1}} \right) \hfill \\[0.5em]
  		2{\xi _1}\left( {{\xi _2}{\xi _3}} \right) - 2{\xi _2}\left( {{\xi _1}{\xi _3}} \right) = {\xi _3}\left( {{\xi _2}{\xi _1}} \right) + {\Gamma ^m}{\xi _3}\left( {{\xi _2}{\Gamma _m}{\xi _1}} \right) \hfill \\ 
		\end{gathered}  \right.
		\label{Fierz}
	\end{equation}

	The Fierz-identities implies several useful formula. Let $\xi_I$ be a symplectic-Majorana spinor and $(s, \kappa, R, \Theta)$ be the associated quantities described in the main text. Then
	\begin{equation}
		\left\{ \begin{gathered}
  		{R^m}{\Gamma _m}{\xi _I} =  - {\xi _I} \hfill \\[0.5em]
  		\Omega _{mn}^ - {\Gamma ^{mn}}{\xi _I} = 0 \hfill \\ 
		\end{gathered}  \right.,\;\;\;\;{t^{IJ}}{\left( {{\Theta _{IJ}}} \right)^m}_k{t^{KL}}{\left( {{\Theta _{KL}}} \right)^k}_n = \frac{{{s^2}}}{2}\left( {{t_I}^J{t_J}^I} \right)\left( { - \delta _n^m + {R^m}{\kappa _n}} \right),
	\end{equation}
for any symmetric tensor $t_{IJ}$ and anti-self-dual (w.r.t to $R^m$) 2-form $\Omega^+$. In particular, if $t_{IJ} \ne 0$ everywhere and satisfies $\overline{t_{IJ}} = \epsilon^{II'}\epsilon^{JJ'} t_{I'J'}$, then the 2-form $t^{IJ} \Theta_{IJ}$ is nowhere-vanishing, since it squares to
	\begin{equation}
		{\left( {{t^{IJ}}{\Theta _{IJ}}} \right)_{mn}}{\left( {{t^{IJ}}{\Theta _{IJ}}} \right)^{mn}} =  - 2{s^2}\left( {{t^{IJ}}{t_{IJ}}} \right) > 0.
	\end{equation}

	\section{Conventions in Differential Geometry}

	In this section we review our convention in differential forms, spin connection and more tensor analysis.

	\underline{\emph{Differential forms}}

	For any differential $p$-form $\omega$, the components ${\omega _{{m_1}...{m_p}}}$ and ${\omega _{{A_1}...{A_p}}}$ are defined as
	\begin{equation}
		\omega  = \frac{1}{{p!}}{\omega _{{m_1}...{m_p}}}d{x^{{m_1}}} \wedge ... \wedge d{x^{{m_p}}} = \frac{1}{{p!}}{\omega _{{A_1}...{A_p}}}{e^{{A_1}}} \wedge ... \wedge {e^{{A_p}}}
	\end{equation}
for coordinate $\{x^m\}$ and vielbein $\{e^A\}$. The wedge product is defined such that
	\begin{equation}
		d{x^m} \wedge d{x^n}\left( {X,Y} \right) = {X^m}{Y^n} - {X^n}{Y^m}
	\end{equation}
The exterior derivative $d$ acting on $\omega$ is then
	\begin{equation}
		d\omega  = \frac{1}{{p!}}{\partial _k}{\omega _{{m_1}...{m_p}}}d{x^k} \wedge d{x^{{m_1}}} \wedge ... \wedge d{x^{{m_p}}}
	\end{equation}
and therefore ${\left( {d\omega } \right)_{k{m_1}...{m_p}}} = \left( {p + 1} \right){\partial _{[k}}{\omega _{{m_1}...{m_p}]}}$. In particular,
	\begin{equation}
		{\left( {d\kappa } \right)_{mn}} = {\partial _m}{\kappa _n} - {\partial _n}{\kappa _m} = \nabla _m^{{\text{LC}}}{\kappa _n} - \nabla _n^{{\text{LC}}}{\kappa _m}
	\end{equation}

	\underline{\emph{Connections}}

	Let $\nabla$ be an arbitrary connection on $TM$, then for any vector $X = X^m \partial_m$, one defines the connection coefficients ${\Gamma^k}_{mn}$ as ${\nabla _m}{X^k} = {\partial _m}{X^k} + {\Gamma ^k}_{mn}{X^n}$. The torsion tensor of such a connection is defined as ${T^k}_{mn} \equiv {\Gamma ^k}_{mn} - {\Gamma ^k}_{nm}$.

	Let $\{e^A\}$ be an orthonormal basis with respect to metric $g$. Then given any connection $\nabla$ preserving $g$, one can write down Cartan structure equation and so define connection 1-form (also called spin connection) ${\omega^A}_B$
	\begin{equation}
		d{e^A} + {\omega ^A}_B \wedge {e^B} = {T^A} \Leftrightarrow {\nabla _m}{e_B} = {\omega _m}^A{_B}{e_A}
	\end{equation}
Preserving the metric $g$ implies anti-symmetric property ${\omega ^A}_B + {\omega ^B}_A = 0$. ${\omega^A}_B$ can be solved from the structure equation, and expressed in terms of ${\Gamma^k}_{mn}$
	\begin{equation}
		{\omega _m}^A{_B} = e_k^Ae_B^n{\Gamma ^k}_{mn} - e_B^n{\partial _m}e_n^A
	\end{equation}

	It is easy to solve the spin connection for the Levi-Civita connection $\nabla^\text{LC}$ of $g$. Suppose $d{e^A} = {C^A}_{BC}{e^B} \wedge {e^C}$ with ${C^A}_{BC} + {C^A}_{CB} = 0$, and ${\omega ^A}_B = {\omega _C}^A{_B}{e^C}$, then
	\begin{equation}
		{\omega _C}^A{_B} =  - {C^A}_{CB} - {C^B}_{AC} + {C^C}_{BA}
	\end{equation}
One can use this to obtain ${\omega _m}^A{_B}{\Gamma ^{AB}}$, or one can exploit the identification
	\begin{equation}
		\sum\limits_{A,B} {{\omega _m}{{^A}_B}{\Gamma ^{AB}}}  \leftrightarrow \sum\limits_{A,B} {{\omega _m}{{^A}_B}{e^A} \wedge {e^B}} 
	\end{equation}
to simplify computation:
	\begin{equation}
		\begin{gathered}
  		\;\;\;\;\;d{e^A} + \sum\limits_B {{\omega ^A}_B{e^B}}  = 0 \Leftrightarrow {\iota _{{\partial _m}}}d{e^A} + \sum\limits_B {{\omega _m}{{^A}_B}{e^B} - e_m^B{\omega ^A}_B}  = 0 \hfill \\
   		\Rightarrow \sum\limits_{A,B} {{\omega _m}{{^A}_B}{e^A} \wedge {e^B}}  =  - \sum\limits_A {\left( {{e^A} \wedge {\iota _{{\partial _m}}}d{e^A} + e_m^Ad{e^A}} \right)}  \hfill \\ 
		\end{gathered} 
	\end{equation}

	Given any connection $\nabla$ that preserves metric $g$, maybe with torsion, one can induce a connection on the spin bundle $S$
	\begin{equation}
		{\nabla _m}\psi  = {\partial _m}\psi  + \frac{1}{4}{\omega _m}^A{_B}{\Gamma ^{AB}}\psi 
	\end{equation}

	We will sometimes use $\cdot$ to denote Clifford action of any differential $p$-form $\omega$ on spinors:
	\begin{equation}
		\omega  \cdot \psi  = \frac{1}{{p!}}{\omega _{{A_1}...{A_p}}}{\Gamma ^{{A_1}...{A_p}}}\psi.
	\end{equation}
So in particular, $\displaystyle d\kappa  \cdot \psi  = \frac{1}{2}d{\kappa _{mn}}{\Gamma ^{mn}}\psi$.

	\underline{\emph{Lie derivative}}

	Let $X$ be a smooth vector field. Then the Lie-derivative $\mathcal{L}_X$ on a differential form $\omega$ is expressed in terms of Cartan identity
	\begin{equation}
		{\mathcal{L}_X}\omega  = {\iota _X}d\omega  + d{\iota _X}\omega 
	\end{equation}
When acting on another vector field $Y$, ${\mathcal{L}_X}Y = \left[ {X,Y} \right]$.

	\section{K-contact Geometry\label{appendix-c}}

	In this appendix we review some basics aspects about contact geometry and K-contact structures. Interested readers may refer to \cite{Blair} for more detail \footnote{However we point out that the convention of exterior derivative $d$ in \cite{Blair} is such that, for instance,\begin{equation}
		d\kappa  = \frac{1}{2}\left( {{\partial _m}{\kappa _n}} \right)d{x^m} \wedge d{x^n}
	\end{equation}}.

	Symplectic geometry is a well-known type of geometry in even dimensions. There, a symplectic structure is defined to be a closed and non-degenerate 2-form $\omega$. In odd dimensions, there is a similar type of structures, called contact structures, which have many similar and interesting behaviors as symplectic structures. 

	\vspace{20pt}
	\underline{\emph{Contact Structure}}

	Let $M$ be a $2n+1$-dimensional compact smooth manifold. The Euler number $\chi(M) = 0$ implies that generic vector fields or 1-forms on $M$ have no zeros. So let $\kappa$ be a nowhere-vanishing 1-form. Then it defines the horizontal vector bundle $TM_H \subset TM$, as we mentioned in the  section \ref{section-1.1}.
	
	In particular, $\kappa$ defines a \emph{contact structure}, or {\emph{contact distribution}} $TM_H$, if it satisfies
	\begin{equation}
		\kappa \wedge (d\kappa)^n \ne 0,\;\;\;\;\text{Everywhere on } M.
		\label{contact-condition}
	\end{equation}
$\kappa$ itself is called a \emph{contact} 1-form of the structure. So in odd dimensions, $d\kappa$ plays the role of symplectic form in even dimensions; indeed, it renders the horizontal bundle $TM_H$ as a symplectic vector bundle of rank $2n$. 

	Once a contact 1-form is given, there is unique vector field $R$ such that
	\begin{equation}
		\kappa_m R^m = 1, \;\;\;\; R^m (d\kappa)_{mn} = 0.
	\end{equation}
and we call it the \emph{Reeb vector field} associated to contact the 1-form $\kappa$. The Reeb vector field on a compact contact manifold generates 1-parameter family of diffeomorphisms (an effective smooth $\mathbb{R}$-action on $M$), which is usually called the \emph{Reeb flow} $\varphi_R(t)$, or the contact flow; the flow translates points along the integral curves of the $R$. It follows from the definition that the flow preserves the contact structure, since $\mathcal{L}_R = \iota_R d\kappa + d \iota_R$ and $\mathcal{L}_R \kappa = 0, \;\mathcal{L}_R d\kappa = 0$.

	It is important to note that the integral curves (or equivalently, the Reeb flow) of $R$ have three types of behaviors:

	1) The {\emph {regular}} type is that all the curves are closed and the Reeb flow generates \emph{free} $U(1)$-action on $M$, rendering $M$ a principal $U(1)$-bundle over some $2n$-dimensional symplectic manifold. 

	2) A \emph{quasi-regular} type is that although the curves are all closed, the Reeb flow only generates \emph{locally-free} $U(1)$-action.

	3) The \emph{irregular} type captures the generic situations, where not all the integral curves are closed. Irregular Reeb flows can have very bad behaviors, but if the Reeb vector field preserves some metric on $M$, then the behavior could still be tractable. In other context, irregular Reeb flows are better than the other two types, in the sense that they are non-degenerate and may provide isolated closed Reeb orbits.
	
	\vspace{20pt}
	\underline{\emph{Contact metric structure}}

	Just as in symplectic geometry, one would like to have some metric and almost conplex structure into the play, so that the contact structure has more ``visible'' properties.
	
	Given a contact 1-form $\kappa$, one can define a set of quantities $(\kappa, R, g, \Phi)$ where $g$ is a metric and $\Phi$ is a $(1,1)$-type tensor, such that
	\begin{equation}
		{g_{mn}}{R^n} = {\kappa _m},\;\;\;\;2 {g_{mk}}{\Phi ^k}_n = {\left( {d\kappa } \right)_{mn}} = {{ \nabla^\text{LC}_m }}{\kappa _n} - {{ \nabla^\text{LC} _n}}{\kappa _m},\;\;\;\;\Phi^2 = -1 + R \otimes \kappa.
	\end{equation}
where $ \nabla^\text{LC} $ denotes the Levi-Civita connection of $g$. We call such set of quantities a \emph{contact metric structure}.
	
	There are a few useful algebraic and differential relations between quantities. First we have
	\begin{equation}
		{\Phi ^n}_m{R^m} = {\kappa _n}{\Phi ^n}_m = 0,\;\;\;\;\frac{{{{\left( { - 1} \right)}^n}}}{{{2^n}n!}}\kappa  \wedge {\left( {d\kappa } \right)^n} = {\Omega _g}.
	\end{equation}
where $\Omega_g$ is the volume form associated to metric $g$. From this one can show that $d\kappa$ satisfies
	\begin{equation}
		{\iota _R}*d\kappa  = d\kappa.
	\end{equation}
And in fact, if one takes an adapted vielbein, for instance in 5-dimension, satisfying ${e_5} = R,\;\;\Phi \left( {{e_{2i - 1}}} \right) = {e_{2i}},\;\;\kappa \left( {{e_{1,2,3,4}}} \right) = 0,\;\;\;\;i = 1,2$, one has
	\begin{equation}
		d\kappa  = 2\left( {{e^1} \wedge {e^2} + {e^3} \wedge {e^4}} \right).
	\end{equation}
	
	Moreover, using ${\iota _R}d\kappa  = 0$ and $\kappa(R) = 1$, it can shown that
	\begin{equation}
		{R^n}{\nabla _m}{\kappa _n} = {\kappa _n}{\nabla _m}{R^n} = {R^m}{\nabla _m}{R^n} = 0,
	\end{equation}
namely $R$ is geodesic.	

	There are useful relations between $R$ and $\Phi$: for any contact metric structure,
	\begin{equation}
		R^m \nabla^\text{LC}_m {\Phi^n}_k = 0.
		\label{Blair_1}
	\end{equation}
and also
	\begin{equation}
		{ \nabla^\text{LC} _m}{R^n} =  - {\Phi ^n}_m - \frac{1}{2}{\left( {\Phi \circ {\mathcal{L}_R}\Phi } \right)^n}_m.
		\label{Blair_2}
	\end{equation}

	\underline{\emph{K-contact structure}}
	
	As we have mentioned earlier, irregular Reeb flows can be more tractable if certain metric is invariant under the flow. This leads to the notion of K-contact structure, where the Reeb vector field is Killing with respect to the metric in a contact metric structure:

	It is called a \emph{K-contact structure}, if a contact metric structure satisfies an additional condition
	\begin{equation}
		\mathcal{L}_R g = 0.
	\end{equation}
Note that this is equivalent to, since $\Phi$ and $d\kappa$ are related by metric $g$, it is easy to see that ${\mathcal{L}_R}\Phi  = 0$, and consequently, ${ \nabla _m}{R^n} =  - {\Phi ^n}_m$.

	\vspace{20pt}
	\underline{\emph{Sasakian Structure}}

	A Sasakian structure is a K-contact structure $(\kappa, R, g, \Phi)$ with additional constraint
	\begin{equation}
		\left( {{\nabla _X}\Phi } \right)Y = g\left( {X,Y} \right)R - \kappa \left( Y \right)X
	\end{equation}
Sasakian structures are K\"{a}hler structures in the odd dimensional world. Therefore, it enjoys many simple properties that allow simplification in computations.

	\vspace{20pt}
	\underline{\emph{Generalized Tanaka-Webster connection}}

	There have been several special connections on contact metric structures introduced in various literatures. For us, the most important one is the generalized Tanaka-Webster connection. There are actually two special connections, both of which are called generalized Tanaka-Webster connection, one introduced by Tanno \cite{Tanno:1989} and the other introduced in \cite{Nicolaescu:aa}. Their names comes from the property that when restricted on a integrable CR structure, the two connections reduces to the usual Tanaka-Webster connection.

	On a general contact metric structure, the two connections are different. However, when the structure is K-contact, the two connections induces the same Dirac operator on the spin bundle $S$ via the standard formula
	\begin{equation}
		{\slashed{\nabla}^{{\text{TW}}}} \equiv {\Gamma ^m}\nabla _m^{{\text{TW}}} = {\Gamma ^m}\left( {{\partial _m} + \frac{1}{4}{{\left( {\omega _m^{{\text{TW}}}} \right)}^A}_B{\Gamma ^{AB}}} \right).
	\end{equation}
In terms of the Levi-Civita connection $\nabla^\text{LC}$, this Dirac operator reads
	\begin{equation}
		{\slashed{\nabla}^{{\text{TW}}}} \psi = {\slashed{\nabla}^{{\text{LC}}}} \psi + \frac{1}{4}d\kappa  \cdot \psi,
	\end{equation}
which is the operator that appears in the localization locus (\ref{localization-locus}). Using the projection $P_{\pm}$ to chiral and anti-chiral spinors, one has for chiral spinor $\forall \phi_+ \in \Gamma(S_+)$
	\begin{equation}
		{P_ - }{\slashed{\nabla} ^{{\text{TW}}}}{\phi _ + } = {P_ - }{\slashed{\nabla} ^{{\text{LC}}}}{\phi _ + },\;\;\;\;{P_+ }{\slashed{\nabla} ^{{\text{TW}}}}{\phi_ + } =  - \left( {\nabla _R^{{\text{LC}}} + \frac{1}{4}d\kappa  \cdot } \right){\phi _ + } =  - \nabla _R^{{\text{TW}}}{\phi _ + }.
	\end{equation}

	\section{$\operatorname{Spin}^\mathbb{C}$ bundle and the Dolbeault-Dirac operator \label{appendix-d}}

	In this appendix we review the $\operatorname{Spin}^\mathbb{C}$ bundles on a contact metric manifold and a canonical Dirac operator on any K-contact structure.

	Consider a contact metric structure $(\kappa, R, g, \Phi)$. Then on the horizontal tangent bundle $TM_H$, $\Phi$ defines a complex structure and thus induces a $(p,q)$-decomposition of the complexification
	\begin{equation}
		T{M_H} \otimes \mathbb{C} = {T^{1,0}}M \oplus {T^{0,1}}M,\;\;\;\;\;\;\;\;{ \wedge ^ \bullet }TM_H^* \otimes \mathbb{C} =  \oplus {T^{p,q}}{M^*}
	\end{equation}

	Let us focus on a 5-dimensional contact metric structure $(M; \kappa, R, g, \Phi)$. One can start from an adapted vielbein $e^A$ as discussed before, and consider the complexification 
	\begin{equation}
		{e^{{z_1}}} \equiv {e^1} + i{e^2},\;\;\;\;{e^{{z_2}}} \equiv {e^3} + i{e^4}.
	\end{equation}
With this complex basis, one sees that $d\kappa$ is of type-$(1,1)$ as expected
	\begin{equation}
		d\kappa  = i\left( {{e^{{z_1}}} \wedge {e^{{{\bar z}_1}}} + {e^{{z_2}}} \wedge {e^{{{\bar z}_2}}}} \right).
	\end{equation}

	The bundle $W^0 \equiv {T^{0, \bullet }}{M^*}$ is also a $\operatorname{Spin}^\mathbb{C}$ bundle in the sense that $TM^*$ acts on it in a Clifford manner
	\begin{equation}
		\left\{ \begin{gathered}
  		\omega  \cdot \psi  = \sqrt 2 i\left( {{\omega _{\bar i}}{{\bar e}^{\bar i}} \wedge \psi  - {g^{\bar ij}}{\omega_j}{\iota _{{e_{\bar i}}}}\psi } \right),\;\;\;\;\omega  = {\omega _i}{e^i} + {\omega _{\bar i}}{{\bar e}^{\bar i}} \in \Gamma \left( {TM_H^*} \right) \hfill \\
  		\kappa \cdot \psi  = {e^1} \cdot {e^2} \cdot {e^3} \cdot {e^4} \cdot \psi  \hfill \\ 
		\end{gathered}  \right..
	\end{equation}
which satisfies the Clifford algebra $\left\{ {\omega  \cdot ,\mu  \cdot } \right\} = 2g\left( {\omega ,\mu } \right)$. In particular, $W^0$ decomposes into chiral and anti-chiral spinor bundle according to the eigenvalue $\pm 1$ of $\Gamma_C \equiv - \kappa \; \cdot$
	\begin{equation}
		{W^0} = W_ + ^0 \oplus W_ - ^0,\;\;\;\;W_ + ^0 \equiv {T^{0,0}}{M^*} \oplus {T^{0,2}}{M^*},\;\;\;\;W_ - ^0 \equiv {T^{0,1}}{M^*}.
	\end{equation}
Using the complex basis ${e^{{\bar z_i}}}$, one can define an orthonormal basis of $W^0$:
	\begin{equation}
		W_ + ^0 = {\text{span}}\left\{ {1,\frac{1}{2}{e^{{{\bar z}_1}}} \wedge {e^{{{\bar z}_2}}}} \right\},\;\;\;\;W_ - ^0 = {\text{span}}\left\{ {\frac{1}{{\sqrt 2 }}{e^{{{\bar z}_1}}},\frac{1}{{\sqrt 2 }}{e^{{{\bar z}_2}}}} \right\}
	\end{equation}
If one represents
	\begin{equation}
		\phi  = {a_1} + \frac{{{a_2}}}{2}{e^{{{\bar z}_1}}} \wedge {e^{{{\bar z}_2}}} + \frac{{{a_3}}}{{\sqrt 2 }}{e^{{{\bar z}_1}}} + \frac{{{a_4}}}{{\sqrt 2 }}{e^{{{\bar z}_2}}} \leftrightarrow {\left( {{a_1},{a_2},{a_3},{a_4}} \right)^T},
	\end{equation}
then the above Clifford action is represented as (\ref{gamma-rep}).

	On a contact metric structure, there may be other $\operatorname{Spin}^\mathbb{C}$ bundles. They can be obtained by tensoring an arbitrary complex line bundle $E$:
	\begin{equation}
		W = {W^0} \otimes E,\;\;\;\;{W_ \pm } = W_ \pm ^0 \otimes E
	\end{equation}
In particular, when the manifold is spin, the spin bundle $S$ can be obtained by
	\begin{equation}
		S = {W^0} \otimes K_M^{ - 1/2} \Leftrightarrow {W^0} = S \otimes K_M^{1/2}
	\end{equation}
where ${K_M} \equiv {T^{0,2}}{M^*}$. More generally, any $\operatorname{Spin}^\mathbb{C}$ bundle $W$ can be written as $W = S \otimes L^{1/2}$ for some complex line bundle $L^{1/2}$ (and its square $L$ is called the determinant line bundle of $W$). For instance, ${W^0} = S \otimes K_M^{1/2}$ and therefore the determinant line bundle $L^0$ of $W^0$ is ${L^0} = {K_M}$. Generally, the determinant line bundle $L$ of $W = {W^0} \otimes E$ is $L = K_M \otimes E^2$.

	This implies that given a connection on $S$ (which can be induced from a metric connection ${\omega ^A}_B$) and a $U(1)$-connection\footnote{A local basis $\sigma$ on $L^{1/2}$ is assumed, such that ${\nabla _A}\left( {f\sigma } \right) = df \otimes \sigma  - iA \otimes \left( {f\sigma } \right)$} $A$ on $L^{1/2}$, we have a connection on $W = S \otimes L^{1/2}$
	\begin{equation}
		{\nabla _A}\psi  = \nabla \psi  - iA\psi ,\;\;\;\;\forall \psi  \in \Gamma (W)
	\end{equation}
The situation of $W^0$ is a bit special, since one can induce a canonical $U(1)$-connection $A_0$ on $K_M$ using the Chern connection $\nabla^\text{C}$ on the almost-hermitian cone $C(M)$. Therefore, taking $A_0$ as a reference connection, any connection $A$ on a $\operatorname{Spin}^\mathbb{C}$ bundle $W$ can be written in terms of  a $U(1)$-connection $a$ on $E$ as $A = \frac{1}{2}{A_0} + a$.

	The above construction is good for any contact metric structure. Now let us focus on a K-contact structure, and use the generalized Tanaka-Webster connection to induce a connection $\nabla^\text{TW}$ on $S$. Combining with a $U(1)$-connection $A$ on $L^{1/2}$, one can define a Dirac operator $\slashed{D}_A^\text{TW}$ \cite{Nicolaescu:aa, Degirmenci:2013aa, Tanno:1989}
	\begin{equation}
		\slashed{D}_A^{{\text{TW}}} \equiv \Gamma  \cdot \nabla _A^{{\text{TW}}}
	\end{equation}

	In \cite{Nicolaescu:aa}, it is shown that when $E$ is trivial and $a = 0$, namely $A = 1/2 A_0$,
	\begin{equation}
		\slashed{D}_{{A_0}/2}^{{\text{TW}}}\left( {\alpha  + \beta } \right) = {\mathcal{L}_R}\left( {\alpha  + \beta } \right) + \bar \partial \alpha  + {\bar \partial ^*}\beta ,\;\;\;\; \alpha + \beta \in {\Omega ^{0,0}} \oplus {\Omega ^{0,2}} = \Gamma(W^0_+)
	\end{equation}
where the Dolbeault operator $\partial$ and $\bar \partial$ are defined in the usual way\footnote{On a K-contact structure, on has in general (recall that $\mathcal{L}_R$ preserves $\Phi$ and therefore the $(p,q)$-decomposition)
	\begin{equation}
		d:{T^{p,q}}{M^*} \to \kappa  \wedge {T^{p,q}}{M^*} \oplus \left( {{T^{p + 1,q}}{M^*} \oplus {T^{p,q + 1}}{M^*} \oplus {T^{p + 2,q - 1}}{M^*} \oplus {T^{p - 1,q + 1}}{M^*}} \right)
	\end{equation}
}
	\begin{equation}
		\partial  \equiv {\pi ^{p + 1,q}}\circ d:{T^{p,q}}{M^*} \to {T^{p + 1,q}}{M^*},\;\;\;\;\bar \partial  \equiv {\pi ^{p,q + 1}}\circ d:{T^{p,q}}{M^*} \to {T^{p,q + 1}}{M^*}
	\end{equation}
Note that the two operators do not square to zero in general; 
define $N\left( {{\omega ^{p,q}}} \right) \equiv {\pi ^{p - 1,q + 2}}\left( {d{\omega ^{p,q}}} \right)$ and $\bar N\left( {{\omega ^{p,q}}} \right) \equiv {\pi ^{p + 2,q - 1}}\left( {d{\omega ^{p,q}}} \right)$, then one has
	\begin{equation}
		{\bar \partial ^2}{\alpha ^{p,q}} =  - N\left( {\partial {\alpha ^{p,q}}} \right) - \partial N\left( {{\alpha ^{p,q}}} \right),\;\;\;\;{\partial ^2}{\alpha ^{p,q}} =  - \bar N\left( {\bar \partial {\alpha ^{p,q}}} \right) - \bar \partial \bar N\left( {{\alpha ^{p,q}}} \right),
	\end{equation}
	\begin{equation}
		\left\{ {\partial ,\bar \partial } \right\}{\omega ^{p,q}} =  - d\kappa  \wedge {\mathcal{L}_R}{\omega ^{p,q}} - \left\{ {N,\bar N} \right\}\left( {{\omega ^{p,q}}} \right),
	\end{equation}
which are almost identical to those on symplectic 4-manifolds, except for the term $d\kappa \wedge \mathcal{L}_R$. On a Sasakian structure, the Nijenhuis map $N$ and $\bar N$ vanish and $\partial^2 = \bar \partial^2 = 0$, similar to K\"{a}hler structure.

	\vspace{20pt}
	\underline{\emph{Weitzenb\"{o}ck Formula}}

	We review several useful formula for studying 5d Seiberg-Witten equations, which are direct generalization from those on symplectic 4-manifolds.

	Consider $W = W^0 \otimes E$ with $U(1)$-connection $a$ on $E$, with curvature $F_a = da$. Then for $ \alpha \in \Omega^{0,0}(E)$, $\beta \in \Omega^{0,2}(E)$, one has Weitzenb\"{o}ck formula
	\begin{equation}
		2\bar \partial _a^*{\bar \partial _a}\alpha  = d_a^{J*}d_a^J\alpha  - \Lambda F_a^{1,1}\alpha  + 2i\mathcal{L}_R^a\alpha ,\;\;\;\;2{\bar \partial _a}\bar \partial _a^*\beta  = \nabla _{{A_0} + a}^*{\nabla _{{A_0} + a}}\beta  - \Lambda {F_{{A_0} + a}} + 2i\mathcal{L}_R^a\beta .
	\end{equation}
where we define operator $d_a^J \equiv {\partial _a} + {{\bar \partial }}_a$, $\nabla_{A_0 + a}$ is the connection induced by $A_0$ and $a$ on $K_M \otimes E$, $\Lambda$ as the adjoint of wedging $d\kappa$:
	\begin{equation}
		\left\langle {{\alpha ^{p - 1,q - 1}},\Lambda {\beta ^{p,q}}} \right\rangle  = \frac{1}{2}\left\langle {d\kappa  \wedge {\alpha ^{p - 1,q - 1}},{\beta ^{p,q}}} \right\rangle , \;\;\;\;\left\langle {\alpha ,\beta } \right\rangle  \equiv \int_M {\alpha  \wedge {*_\mathbb{C}}\beta } 
	\end{equation}
The Weitzenb\"{o}ck formula can be shown using K\"{a}hler identities
	\begin{equation}
		i\bar \partial _a^*{\omega ^{p,q}} = \left[ {\Lambda ,{\partial _a}} \right]{\omega ^{p,q}},\;\;\;\; - i\partial _a^*{\omega ^{p,q}} = \left[ {\Lambda ,{{\bar \partial }_a}} \right]{\omega ^{p,q}}, \;\;\;\; \forall \omega^{p,q} \in \Omega^{p,q}(E).
	\end{equation}
and the fact that the Dolbeault operators can be expressed in terms of $\nabla^\text{TW}$
	\begin{equation}
		\bar \partial  = {e^{{{\bar z}_i}}} \wedge {\nabla^\text{TW} _{{e_{{z_i}}}}},\;\;\;\;{\bar \partial ^*} =  - 2\iota \left( {{e_{{{\bar z}_i}}}} \right){\nabla^\text{TW} _{{e_{{z_i}}}}}.
	\end{equation}
for an adapted complex vielbein.

\end{appendices}

\bibliographystyle{JHEP}
\bibliography{References}
\end{document}